\documentclass[aps,prb,twocolumn,groupedaddress,showpacs]{revtex4}
\usepackage{graphicx}
\usepackage{dcolumn}
\usepackage{bm}
\usepackage{amsmath,amsfonts,amssymb,amsthm,verbatim}
\usepackage[ansinew]{inputenc}
\usepackage[usenames]{pstcol}

\usepackage{color}
\usepackage{epsfig}

\newcommand{\be}{\begin{equation}}
\newcommand{\ee}{\end{equation}}
\newcommand{\bea}{\begin{eqnarray}}
\newcommand{\eea}{\end{eqnarray}}

\newcommand{\bi}{\begin{itemize}}
\newcommand{\ei}{\end{itemize}}
\newcommand{\bc}{\begin{center}}
\newcommand{\ec}{\end{center}}

\newcommand{\la}{\langle}
\newcommand{\ra}{\rangle}

\newcommand{\p}{\partial}

\newcommand{\s}{\sigma}
\newcommand{\reff}{\text{eff}}

\begin{document}

\title{Modulated Rashba interaction in a quantum wire: Spin and charge dynamics}
\author{Mariana Malard}
\affiliation{Faculdade UnB Planaltina, Universidade de Brasilia,
73300-000 Planaltina-DF, Brazil}
\author{Inna Grusha and G. I. Japaridze}
\affiliation{Andronikashvili Institute of Physics, Tamarashvili 6,
0177 Tbilisi, Georgia and\\
College of Engineering, Ilia State University, Cholokasvili Ave.
3-5, 0162 Tbilisi, Georgia}
\author{Henrik Johannesson}
\affiliation{Department of Physics, University of Gothenburg, SE 412
96 Gothenburg, Sweden}

\begin{abstract}

It was recently shown that a spatially modulated Rashba spin-orbit
coupling in a quantum wire drives a transition from a metallic to an
insulating state when the wave number of the modulation becomes
commensurate with the Fermi wave length of the electrons in the wire
[G. I. Japaridze {\em et al.}, Phys. Rev. B {\bf 80} 041308(R)
(2009)]. On basis of experimental data from a gated InAs
heterostructure it was suggested that the effect may be put to
practical use in a future spin transistor design. In the present
article we revisit the problem and present a detailed analysis of
the underlying physics. First, we explore how the build-up of charge
density wave correlations in the quantum wire due to the periodic
gate configuration that produces the Rashba modulation influences
the transition to the insulating state. The interplay between the
modulations of the charge density and that of the spin-orbit
coupling turns out to be quite subtle: Depending on the relative
phase between the two modulations, the joint action of the Rashba
interaction and charge density wave correlations may either enhance
or reduce the Rashba current blockade effect. Secondly, we inquire
about the role of the Dresselhaus spin-orbit coupling that is
generically present in a quantum wire embedded in semiconductor
heterostructure. While the Dresselhaus coupling is found to work
against the current blockade of the insulating state, the effect is
small in most materials. Using an effective field theory approach,
we also carry out an analysis of effects from electron-electron
interactions, and show how the single-particle gap in the insulating
state can be extracted from the more easily accessible collective
charge and spin excitation thresholds. The smallness of the
single-particle gap together with the anti-phase relation between
the Rashba and chemical potential modulations pose serious
difficulties for realizing a Rashba-controlled current switch in an
InAs-based device. Some alternative designs are discussed.

\end{abstract}


\maketitle

\begin{center}
\underline{Published version}

\vspace{.5cm}

Phys. Rev. B {\bf 84}, 075466 (2011).

DOI: 10.1103/PhysRevB.84.075466

\end{center}

\section{Introduction}

The ability to control and manipulate electron spins in
semiconductors via an external electric field forms the basis of the
emerging spintronics technology \cite{Review}. In what has become a
paradigm for the next-generation spintronics device - the Datta-Das
spin field effect transistor \cite{DattaDas} - spin-polarized
electrons are injected from a ferromagnetic emitter into a quantum
wire patterned in a semiconductor heterostructure. The Rashba
spin-orbit interaction \cite{Rashba} intrinsic to a quantum well
patterned in a semiconductor heterostructure causes spin flips of
the injected electrons with a rate tunable by an electrical gate,
and by contacting a ferromagnetic collector to the other end of the
wire, electrons are either accepted or rejected depending on their
spin directions. However, present techniques for injecting
spin-polarized electrons from a ferromagnetic metal into a
semiconductor are quite inefficient. This, among other difficulties,
has obstructed the actual fabrication of a Datta-Das transistor. The
best efficiency rates to date, using a Schottky contact for spin
injection, are still far below what is required for a working device
\cite{FabianZutic}. While other designs for spin transistors have
been proposed, these suffer from similar technical difficulties as
the original Datta-Das proposal. Alternative blueprints for spin
transistors that do not rely on spin-polarized electron injection
are thus very much wanted.

In a recent work it was shown that a smoothly modulated Rashba
spin-orbit coupling in a quantum wire drives a transition from a
metallic to an insulating state when the wave number of the
modulation becomes commensurate with the Fermi wavelength of the
electrons in the wire \cite{JJF_Paper_09}. It was suggested that
this effect may be put to practical use in a device where a
configuration of equally spaced nanosized gates are placed on top of
a biased quantum wire. When charged, the gate configuration produces
a periodic modulation of the Rashba interaction, thus blocking the
current when the electron density is tuned to commensurability by an
additional backgate. By decharging the gate, the current is free to
flow again. This would realize an ``on-off"  current switch,
controllable by the backgate. The advantage of this proposal is
precisely that it dispenses with the need to inject spin-polarized
electrons into the current-carrying channel of the device.

The proposal in Ref. \onlinecite{JJF_Paper_09} was inspired by
earlier work by Wang \cite{Wang} and Gong and Yang \cite{GongYang},
showing that a current in a quantum wire where segments with a
uniform Rashba coupling alternates with segments with no coupling
gets blocked when the number of segments becomes sufficiently large
\cite{Kirczenow}. However, the Peierls-type mechanism of the
spin-based current switch identified in Ref.
\onlinecite{JJF_Paper_09} is very different from that  in Refs.
\onlinecite{Wang,GongYang}, where the current blockade is simply
caused by electron scattering at the artificially sharp boundaries
between the wire segments (similar to the scattering off the
boundary between the wire and the ferromagnetic collector in the
Datta-Das transistor). Importantly, by instead modeling the Rashba
interaction as {\em smoothly} modulated $-$ thus faithfully taking
into account the fact that the top gates that produce the effective
Rashba field are of finite extent $-$ yields the extra bonus of
allowing for a well-controlled analysis of effects from
electron-electron interactions \cite{JJF_Paper_09}. It was found
that in the experimentally relevant parameter range, the
electron-electron interactions {\em enhance} the current blockade
effect, thus assisting the use of a gate-controlled modulated Rashba
interaction as a current switch.

In the present article we revisit the problem to obtain a more
detailed picture of the underlying physics. First, we shall explore
how the build-up of charge density wave (CDW) correlations in the
quantum wire due to the presence of the periodic gate configuration
influences the current blockade caused by the modulated Rashba
interaction. While one would maybe expect the concurrent modulation
of the charge density to always assist the current blockade, the
interplay between the two effects turns out to be more subtle: When
the two modulations are in phase they do work in tandem, but when
anti-phased a \emph{crossover} regime is observed where the two
modulations compete with each other and, as a result, the Rashba
current blockade effect is reduced by the joint action of the Rashba
interaction and CDW correlations. While at first surprising, we
shall be able to provide a simple explanation of this crossover
effect. Secondly, we shall inquire about the role of the {\em
Dresselhaus spin-orbit interaction}  present in any semiconductor
heterostructure that supports a quantum wire (as most
heterostructures used in experiments are made out of compounds with
broken lattice inversion symmetry, thus implying the presence of a
Dresselhaus interaction) \cite{Dresselhaus}. The Dresselhaus
interaction is found to oppose the current-blockade effect, but as
long as the Rashba interaction dominates that of Dresselhaus, the
effect is small and does not detract from the viability of using a
modulated Rashba interactions as the {\em modus operandi} for a
novel type of spin transistor.

The rest of the paper is organized as follows: In Sec. II we lay the
groundwork and construct the minimal model that captures the effect
of a modulated Rashba spin-orbit interaction in a quantum wire. In
Sec  III we show that a stripped-down version of the model $-$
describing noninteracting electrons $-$ can be mapped onto two
independent sine-Gordon models using bosonization. We perform a
renormalization-group (RG) analysis of the relevant low-energy limit
of the theory, and extract the condition for an opening of a mass
gap in the spin- and charge sectors. In Sec. IV we extend the
analysis to the realistic case of interacting electrons. This
analysis is patterned upon that in the previous section, albeit with
some added technical subtleties. By carrying it out with Sec. III as
a template, we believe that our results will gain in transparency
and ease of interpretation.  Again we extract the condition under
which an insulating gap opens, allowing us to assess the
effectiveness of using a gate-controlled modulated Rashba
interaction as a current switch. In Sec. V we then carry out a case
study, using our results to predict the size of the gap for a
quantum wire patterned in a gated InAs-based heterostructure for
which good experimental data are available. While we find that for
this particular structure the gap will be too small to be usable for
a current switch, our analysis points the way to more effective
designs. Finally, in Sec. VI we summarize our results. Throughout
the paper we try to provide enough detail to make it essentially
self-contained to a reader with some acquaintance with bosonization
and perturbative RG methods.

\section{The model}

In the following we consider a set-up with a 1D quantum wire formed
in a gated 2D quantum well supported by a semiconductor
heterostructure. We assume that the electrons in the wire are
ballistic, restricting us to wire lengths on the micronscale for
most materials. Moreover, by modeling the wire as an {\em ideal} 1D
wire that carries only one conduction channel, we will neglect
effects from the transverse confining potential. This simplification
greatly facilitates our analysis, but, as we shall argue, has little
or no effects on our results. In the standard tight-binding
formalism \cite{FUL}, the kinetic energy and the chemical potential
as well as the interaction energy between the electrons in the wire
are described by the lattice Hamiltonians $H_0$ and $H_{\text{e-e}}$
respectively, with
\begin{eqnarray}
H_0 &\!=\! & -t \sum_{n,\zeta}\! \left(
c^{\dagger}_{n,\zeta}c^{\phantom{\dagger}}_{n+1,\zeta}
\!+\!\mbox{H.c.}\right)  \!-\! \mu\sum_{n, \zeta}c^{\dagger}_{n,\zeta}c^{\phantom{\dagger}}_{n,\zeta}\;, \label{free}\\
H_{\text{e-e}}&\!=\!& \frac{1}{2}\sum_{n,n',\zeta,
\zeta'}V(n-n')c^{\dagger}_{n,\zeta} c^{\dagger}_{n',\zeta'}
c^{\phantom{\dagger}}_{n',\zeta'}c^{\phantom{\dagger}}_{n,\zeta}\;.
\label{e-e}
\end{eqnarray}
Here $c^{\dagger}_{n,\zeta}$ ($c^{\phantom{\dag}}_{n,\zeta}$) is the
creation (annihilation) operator for an electron with spin
${\zeta}=\uparrow,\downarrow$ on site $n$, $t$ is the electron
hopping amplitude, and $\mu$ a uniform chemical potential
controllable by an electrical backgate. The Coulomb interaction
$V(n-n')$ between electrons at sites $n$ and $n'$ is screened by the
metallic gates in the device, with the screening length set by the
distance to the nearest gate \cite{Hausler}.

The electrons in a 2D quantum well are subject to two types of
spin-orbit interactions, the {\em Dresselhaus} \cite{Dresselhaus}
and {\em Rashba} \cite{Rashba} interactions, both originating from
the inversion asymmetry of the potential
$V(\boldsymbol{r})=V_{\text{cr}}(\boldsymbol{r})+V_\text{ext}(\boldsymbol{r})$,
where $V_{\text{cr}}(\boldsymbol{r})$ is the periodic crystal
potential, and $V_\text{ext}(\boldsymbol{r})$ is the aperiodic part
containing effects from other sources (quantum well confinement,
impurities, electrical gates, etc.). The potential gradient  $\nabla
V(\boldsymbol{r})$ produces a Pauli spin-orbit interaction that can
be written as
\begin{equation}
H_{SO} = \lambda_\text{cr}\left( \boldsymbol{k} \times \nabla
V_\text{ext}(\boldsymbol{r}) \right) \cdot \boldsymbol{\sigma} \, -
\, \boldsymbol{b}(\boldsymbol{k}) \cdot \boldsymbol{\sigma},
\label{Pauli}
\end{equation}
where, in the first term, the contribution from $V_\text{cr}$ has
been absorbed in the effective constant $\lambda_\text{cr}$, while
in the second term, $\boldsymbol{b}(\boldsymbol{k})$ is an intrinsic
spin-orbit field  produced by $V_\text{cr}$ only. Here
$\boldsymbol{k}$ is the wave number of an electron, with
$\boldsymbol{\sigma}$ the vector of Pauli matrices representing its
spin. In semiconductors where the crystal potential lacks inversion
symmetry, i.e. $V_\text{cr}(\boldsymbol{-r}) \neq
V_\text{cr}(\boldsymbol{r})$ (including zinc-blende lattice
structures, to which the often used GaAs and InAs quantum wells
belong), the internal spin-orbit field $
\boldsymbol{b}(\boldsymbol{k})$ in Eq. (\ref{Pauli}) fails to
average to zero in a unit cell, resulting in a spin splitting
encoded by the effective Dresselhaus interaction \cite{Dresselhaus}.
For a heterostructure grown along $[001]$, with the electrons
confined to the quantum well in the $xy$-plane, the leading term in
the Dresselhaus interaction takes the simple form
\begin{equation}
H_{\beta} = \beta (k_x \sigma_x - k_y \sigma_y),
\label{Dresselhaus2D}
\end{equation}
with $\beta$ a material- and structure-dependent parameter
\cite{DK}.

The spin degeneracy in a quantum well can be lifted also because of
the structure inversion asymmetry of the confining potential
contained in $V_\text{ext}(\boldsymbol{r})$. More precisely, the
spatial asymmetry of the edge of the conduction band along the
growth direction of the quantum well (i.e. in the $z$-direction
perpendicular to the symmetry plane of the well) mimics an electric
field in that same direction, and one obtains from Eq. (\ref{Pauli})
the Rashba interaction \cite{Rashba}
\begin{equation}
H_{\alpha} = \alpha (k_x \sigma_y - k_y \sigma_x). \label{Rashba2D}
\end{equation}
The Rashba coupling $\alpha$ has a complex dependence on several
distinct features of the quantum well, including the ion
distribution in the nearby doping layers \cite{Sherman}, the
relative asymmetry of the electron density at the two quantum well
interfaces \cite{GolubIvchenko}, and importantly, the applied gate
electric field \cite{REVIEW}. The latter feature allows for a gate
control of the Rashba coupling $\alpha$, with a variation of more
than a factor of two from its base value reported for InAs quantum
wells \cite{Grundler}. One must realize that $\alpha$ in Eq.
(\ref{Rashba2D}) is a spatial average of a microscopic randomly
fluctuating Rashba coupling. In a zinc-blende lattice structure the
fluctuations can be quite large, with a root-mean square deviation
roughly of the same size as the average $\alpha$  \cite{Sherman}. As
discussed in Ref. \onlinecite{SJJ}, for quantum wells with an
anomalously large Rashba coupling $-$ as in the HgTe quantum wells
which support quantum spin Hall states $-$ this large disordering
effect may cause an Anderson transition to an insulating state when
the electron-electron interaction is weakly screened. In other
zinc-blende lattice structures, like GaAs or InAs favored in most
spintronics applications, the disordering effect is weaker, with a
Rashba-induced localization length that is expected to be much
longer than the mean-free path due to impurity scattering. Having
already assumed that the wire has a length that is smaller than the
mean free path, we can therefore ignore the random fluctuations in
the Rashba coupling in what follows.

Projecting the  Dresselhaus and Rashba interactions in Eqs.
(\ref{Dresselhaus2D}) and (\ref{Rashba2D}) along the direction
$\hat{x}$ of the quantum wire and using the same tight-binding
lattice formalism as in Eqs. (\ref{free}) and (\ref{e-e}), one
obtains
\begin{equation} \label{DR}
H_{DR}=-i\!\sum_{n,\zeta,\zeta'}
c^{\dag}_{n,\zeta}\!\left[\gamma_{D}\, \sigma^{x}_{\zeta\zeta'} \!+
\!\gamma_{R}\, \sigma^{y}_{\zeta\zeta'}\right]
\!c^{\phantom{\dag}}_{n\!+\!1,\zeta'} + \mbox{H.c.},
\end{equation}
where $\gamma_D \!=\! \beta a^{-1}$, $\gamma_R \!= \!\alpha a^{-1}$,
with $a$ the lattice spacing. The relative sign and magnitude of
$\gamma_D$ and $\gamma_R$ depends on the material as well as on the
particular design of the heterostructure, with $|\gamma_D| \approx
|\gamma_R| \approx 5 \times 10^{-2}$ meV in a typical GaAs-based
quantum well, while in a HgTe quantum well the Rashba coupling is
orders of magnitude larger than that of Dresselhaus, with
$|\gamma_R| \gtrapprox 10^2 \times |\gamma_D| \approx 10 $ meV
\cite{Silsbee}. Let us mention in passing that the effect of a {\em
uniform} spin-orbit interaction on the electron dynamics in a
quantum wire has been theoretically investigated for both
noninteracting \cite{ NONINTERACTING} and interacting electrons
\cite{INTERACTING,GJPB_05,Egger2009}, and is by now well understood.

\begin{figure}[htpb]
\begin{center}
\includegraphics[scale=0.35]{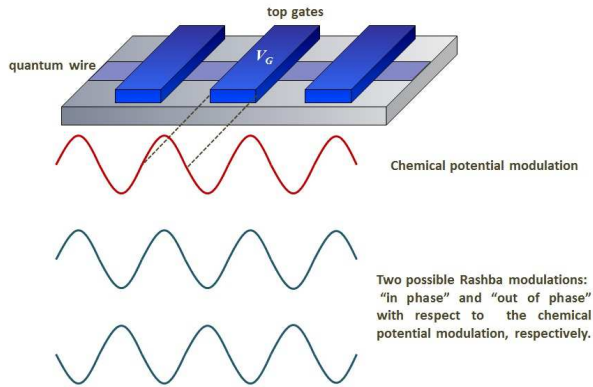}
\caption{(Color online) Schematic figure of the device studied in
the paper. The modulated chemical potential and Rashba interaction
are shown for both ``in phase" and ``out of phase" modulations. The
amplitudes of the chemical potential and Rashba modulations are
proportional to the gate voltage $V_G$.} \label{Fig:fig1}
\end{center}
\end{figure}

We shall assume that the wire is patterned in a heterostructure on
top of which are placed a periodic sequence of equally sized
nanoscale gates, positively charged and pairwise separated by the
same distance as their extensions along the direction of the wire.
The gates may be realized by a series of ultrasmall capacitively
coupled metallic electrodes deposited on the top of the
heterostructure, as illustrated in Fig. \ref{Fig:fig1}. By charging
the gates, one produces a periodic modulation of the Rashba
coupling, together with a concurrent modulation of the local
chemical potential in the wire, with amplitudes depending on the
associated voltage drop across the well (proportional to the gate
voltage $V_G$). The modulation will be smoothly varying along the
wire, reflecting the finite extent of the gates in addition to
effects from distortions and stray electric fields. To a good
approximation, the modulation can be represented by a simple
harmonic and thus we may write
\begin{eqnarray} \label{mod}
H_\text{mod}\!=\!& i\!\!&\!\!\sum_{n,\zeta,\zeta'}\left(
\gamma_{R\text{mod}} \cos(qna) c^{\dag}_{n,\zeta}
\sigma^{y}_{\zeta\zeta'}c_{n+1,\zeta'}\! -\! \mbox{H.c.} \right) \nonumber \\
&-&\sum_{n,\zeta} \mu_{\text{mod}} \cos(qna)
c^{\dag}_{n,\zeta}c_{n,\zeta}\;.
\end{eqnarray}

Here $\mid\!\!\gamma_{R\text{mod}}\!\!\mid \ (\mu_{\text{mod}})$ is
the amplitude of the Rashba field (local chemical potential)
modulation, both of wave number $q$. Note that the Rashba coupling
and the chemical potential modulations are ``in phase" when
$\gamma_{R\text{mod}} > 0$ (and hence has the same sign as
$\mu_{\text{mod}}$, which is always positive), while for
$\gamma_{R\text{mod}} < 0$  the two modulations are ``out of phase"
by $\pi$. The two possible phase relations between the Rashba and
the chemical potential modulations are illustrated in Fig.
\ref{Fig:fig1}. Assuming that the gate electrodes which produce the
chemical potential modulation are positively charged, the segment of
the wire below a gate has an enhanced magnitude of the local
chemical potential (see Fig. \ref{Fig:fig1}), but with a negative
sign. Note that the negative sign has been taken out of the sum in
Eq. (\ref{mod}) (as well as in Eq. (\ref{free}) which contains the
{\em uniform} chemical potential)\cite{ChemPotFootnote}.

The second-quantized expression for the lattice Hamiltonian
$H_{\text{mod}}$ in Eq. (\ref{mod}) provides a microscopic
definition of the Rashba and chemical potential modulations and is
manifestly Hermitian by virtue of the subtraction of the H.c.-term.
This procedure replaces the symmetrization $\alpha(x)
(-i\partial/\partial x) \rightarrow \{\alpha(x), -i\partial/\partial
x \}/2$ for a spatially varying Rashba interaction
$\alpha(x)(-i\partial/\partial x)\sigma_y$ often employed in the
literature as a means to ensure Hermiticity in a first-quantized
continuum formalism \cite{ShermanSinova}.

It is important to stress that the relation between a spatially
modulated gate bias and a Rashba interaction may be more complex
than transpires from our simple model. Already when tuning a gate
voltage that is {\em uniform} along a quantum wire patterned in a
heterostructure, the Rashba interaction has been shown to sometimes
respond in a surprising way, even reversing its sign without a
reversal of the gate bias \cite{Litvinov,MatsudaYoh}. In fact, the
details of how the various effects mentioned above (including the
external gate voltage), influence the magnitude and the sign of the
Rashba parameter in a gated heterostructure have proven notoriously
difficult to sort out, and remains a somewhat contagious issue
\cite{Sandoval}. We shall not attempt to add to this discussion, but
instead focus on the physics implied by the idealized situation
described by $H_{\text{mod}}$ in Eq. (\ref{mod}).

Having defined our model by the Hamiltonian
\begin{equation} \label{model}
H = H_0+ H_{\text{e-e}} + H_{DR} + H_\text{mod},
\end{equation}
with $H_0, H_{\text{e-e}}, H_{DR}$ and $H_\text{mod}$ given by Eqs.
(\ref{free}), (\ref{e-e}),  (\ref{DR}), and (\ref{mod}),
respectively, it is now convenient to pass to a basis which
diagonalizes the uniform spin-orbit interaction $H_{DR}$. For this
purpose we first perform a rotation of the coordinate system by an
angle $2\theta=\arctan(\gamma_D/\gamma_R)$ around the $\hat{z}$-axis
to select the direction of the combined uniform Rashba and
Dresselhaus field, $\sim \gamma_D\, \hat{x} + \gamma_R\, \hat{y}$,
as our new $\hat{y}^{\prime}$-axis:
\begin{equation} \label{rotation}
e^{-i\theta \sigma_z} {\large[}\gamma_D \sigma_x +  \gamma_R
\sigma_y {\large]} e^{i\theta \sigma_z} =
\gamma_\text{eff}\,\sigma_{y^{\prime}},
\end{equation}
where $\gamma_\text{eff} = \sqrt{\gamma_R^2 + \gamma_D^2}$. We then
introduce a spinor basis which diagonalizes $\sigma_{y^{\prime}}$,
\begin{equation} \label{spinor}
 \left( \begin{array}{c}
d_{n,+} \\
d_{n,-} \end{array} \right) \equiv \frac{1}{\sqrt{2}} \left(
\begin{array}{c}
e^{-i\theta}c_{n,\uparrow} - ie^{i\theta}c_{n,\downarrow} \\
-ie^{-i\theta}c_{n,\uparrow} + e^{i\theta}c_{n,\downarrow}
\end{array} \right),
\end{equation}
where the spinor components $\tau\! = \!\pm$ of the operator $d_{n,
\tau}$ label the new quantized spin projections along
$\hat{y}^{\prime}$, with $y^{\prime}$ defining the orientation of
the axis around which the expectation value of the spin will now be
precessing. With this, we write the transformed Hamiltonian as
\begin{equation} \label{transformed}
H^{\prime}= H_0^{\prime}+H_{\text{e-e}}^{\prime} + H_{DR}^{\prime} +
H_\text{mod}^{\prime},
\end{equation}
with
\begin{eqnarray}\label{H_Hubbard_transformed}
H_0^{\prime} \!&\!=\! &\! -t \sum_{n,\tau}\! \left(
d^{\dagger}_{n,\tau}d^{\phantom{\dagger}}_{n+1,\tau}\!+\!\mbox{H.c.}\right)  \!-\! \mu\sum_{n,\tau} d^{\dagger}_{n, \tau}d^{\phantom{\dagger}}_{n, \tau} \label{freeTransformed}  \\
H_{\text{e-e}}^{\prime}\!&\!=\!&\! \frac{1}{2}\sum_{n,n',\tau,
\tau'}\!V(n-n')d^{\dagger}_{n,\tau} d^{\dagger}_{n',\tau'}
d^{\phantom{\dagger}}_{n',\tau'}d^{\phantom{\dagger}}_{n,\tau}
\label{e-eTransformed}
\end{eqnarray}
\begin{eqnarray} \label{H_DR_transformed}
H_{DR}^{\prime} &=& - i\, \gamma_\text{eff} \sum_{n,\tau} \tau
\,d^{\dag}_{n,\tau} d^{\phantom{\dag}}_{n\!+\!1, \tau} +
\mbox{H.c.},
\end{eqnarray}
\begin{eqnarray} \label{H_mod_transformed}
H_\text{mod}^{\prime} = &- & i\sum_{n,\tau} \gamma_R(n) \tau
\cos(2\theta)
d^{\dag}_{n,\tau} d^{\phantom{\dag}}_{n\!+\!1, \tau} \nonumber  \\
&+& i\sum_{n,\tau}  \gamma_R(n) \sin(2\theta) d^{\dag}_{n,\tau} d^{\phantom{\dag}}_{n\!+\!1, -\tau} \nonumber \\
&-&\frac{1}{2} \sum_{n, \tau } \mu(n)
d^{\dagger}_{n,\tau}d^{\phantom{\dagger}}_{n,\tau} + \mbox{H.c.},
\end{eqnarray}
with $\gamma_R(n)\!\equiv\! \gamma_{R\text{mod}} \cos(qna)$ and
$\mu(n)\!\equiv\!  \mu_{\text{mod}} \cos(qna)$.

Let us add a comment that our procedure leading up to Eqs.
(\ref{transformed}) - (\ref{H_mod_transformed}) is not to be
confounded with the gauge transformation approach to two-dimensional
spin-orbit interactions recently suggested by Tokatly and Sherman
\cite{TokatlySherman} (see also Ref. [\onlinecite{LevitovRashba}]).
Whereas our transformation is simply a global spinor rotation, the
gauge transformation in Ref. [\onlinecite{TokatlySherman}] is by
construction a local rotation, yielding a manifest spin-charge
duality. It would be interesting to explore whether the approach by
Tokatly and Sherman\cite{TokatlySherman} can be adapted to the case
also of a modulated spin-orbit interaction, but for now we leave
this for the future.

While the theory defined by Eqs. (\ref{transformed})\, -\,
(\ref{H_mod_transformed}) may look forbiddingly complex, we shall
find that a bosonization approach yields a well-controlled
analytical solution in the physically relevant limit of low
energies. In the next section we study the case with no
electron-electron interaction, i.e. with $V(n-n')=0$ for all $n, n'$
in Eq. (\ref{e-eTransformed}). This simplification allows us to
focus on the key elements of our solution approach, paving the
ground for the more elaborate analysis of the full theory in Sec.
IV.

\section{Non-interacting electrons}

 \subsection{Effective Hamiltonian}

Neglecting the electron-electron interaction $H_{\text{e-e}}$, and
taking $\gamma_R(n) = \mu(n) = 0$ (assuming that there is no
modulated electric field present), the remaining piece of the
Hamiltonian in Eq. (\ref{transformed}), $H_0^{\prime} +
H_{DR}^{\prime}$,  is easily diagonalized by a Fourier transform,
\begin{equation}
H_0^{\prime}+H_{DR}^{\prime}=\sum_{k,\tau =
\pm}\,E^{(0)}_{\tau}(k)d^{\dag}_{k,\tau}d^{\phantom{\dagger}}_{k,\tau}.
\label{H0tildaMS}
\end{equation}
Here
\begin{equation}
E^{(0)}_{\tau}(k)=-2\tilde{t}\cos[(k+\tau q_{0})a]-\mu,
\label{E0tilda}
\end{equation}
with $\tilde{t}=\sqrt{t^{2}+\gamma^2_{\text{eff}}}$ and
$q_{0}a=\arctan(\gamma_{\text{eff}}/t)$, and where $a$ is the
lattice constant. At band-filling $\nu=N_{e}/2N_{0}$, with $N_e \,
[N_0]$ the number of electrons [lattice sites], the system is
characterized by the four Fermi points $k_{F,R}^{\tau} = k_{F}\, +
\tau q_{0}, k_{F,L}^{\tau} =  - k_{F} + \, \tau q_{0} \, \, (\tau =
\pm)$, where $k_{F}= \pi \nu/a$, reflecting the band splitting
caused by the uniform spin-orbit interaction $H^{\prime}_\text{DR}$
in Eq. (\ref{H_DR_transformed}).

To analyze the effect of adding the modulated term
$H_\text{mod}^{\prime}$ in Eq. (\ref{H_mod_transformed}) to
$H_0^{\prime} + H_{DR}^{\prime}$, it is convenient to linearize the
spectrum around these Fermi points and then pass to a continuum
limit with $na \rightarrow x$. By decomposing the lattice operators
$d^{\phantom{\dagger}}_{n,\tau}$ into right- and left-moving fields
$R^{\phantom{\dagger}}_{\tau}(x)$ and
$L^{\phantom{\dagger}}_{\tau}(x)$,
\begin{displaymath}  \label{Decomposition}
d^{\phantom{\dagger}}_{n,\tau}  \rightarrow  \sqrt{a}
\big(\mbox{e}^{ik_{F,R}^{\tau}x} R^{\phantom{\dagger}}_{\tau}(x) +
\mbox{e}^{i k_{F,L}^{\tau}x}L^{\phantom{\dagger}}_{\tau}(x) \big),
\end{displaymath}
we find that in this limit $H_0^{\prime} +
H_{DR}^{\prime}\!+\!H_\text{mod}^{\prime}\! = \!
 \sum_{\tau} \int dx\left({\cal H}_{\tau} + \mbox{H.c.}\right)$, with
\begin{widetext}
 \begin{eqnarray}  \label{linear-Hamiltonian}
{\cal H}_{\tau} \!&=& -i(v_F/2)
\big(\!:\!R^{\dag}_{\tau}(x)\partial_{x}R^{\phantom{\dagger}}_{\tau}(x)\!:-
:\!L^{\dag}_{\tau}(x)\partial_{x}L^{\phantom{\dagger}}_{\tau}(x)\!:\!\big)
-(\lambda_R \mbox{e}^{-i\pi\nu} + \mu_{\text{mod}}) \cos(q
x)\mbox{e}^{-2ik_Fx}
R^{\dagger}_{\tau}(x)L^{\phantom{\dagger}}_{\tau}(x) \nonumber\\
&+&i\lambda_D\sin(\pi\nu)\cos(qx)\mbox{e}^{-iq_0\tau(2x+a)}
\big(R^{\dag}_{\tau}(x)R^{\phantom{\dagger}}_{-\tau}(x)-
L^{\dag}_{\tau}(x)L^{\phantom{\dagger}}_{-\tau}(x)\big).
\end{eqnarray}
\end{widetext}
Here $v_{F}=2a\tilde{t}\sin(\pi\nu)$, $\lambda_{R}= 2{\tilde
\gamma}_{R}\sin(q_{0}a)$, and $\lambda_{D}={\tilde \gamma}_{D}$,
with ${\tilde \gamma}_{j}=
\gamma_{R\text{mod}}\gamma_{j}(\gamma_{R}^{2}+\gamma^{2}_{D})^{-1/2},
\, j=R,D$. The normal ordering $:...:$ is carried out with respect
to the filled Dirac sea. Note that in deriving Eq.
(\ref{linear-Hamiltonian}) we have omitted all rapidly oscillating
terms that vanish upon integration.


The Hamiltonian in Eq. \! (\ref{linear-Hamiltonian}) supports four
distinct limiting cases, depending on the difference  between the
modulation wave number $q$ and the parameters $k_F, q_0$:
\begin{eqnarray}
 (i) & |q \pm 2k_F| \simeq  {\cal O}(1/a), \  |q \pm 2q_0|  \simeq  {\cal O}(1/a);  \nonumber \\
 (ii) &  |q \pm 2k_F|  \simeq  {\cal O}(1/a), \ |q - 2q_0| \ll  {\cal O}(1/a); \nonumber   \\
(iii) &  |q - 2k_{F}|  \ll  {\cal O}(1/a), \  |q \pm 2q_0|  \simeq  {\cal O}(1/a);  \nonumber \\
(iv) &  |q - 2k_{F}|  \ll  {\cal O}(1/a), \  |q-2q_0|  \ll  {\cal
O}(1/a). \nonumber
\end{eqnarray}

In the first case $(i)$, all terms in Eq. (\ref{linear-Hamiltonian})
proportional to $\lambda_R$ or $\mu_{\text{mod}}$ or $\lambda_D$ are
rapidly oscillating and thus average to zero when integrated. It
follows that in this limit the model describes a two-component free
Fermi gas, i.e. a metallic phase with gapless excitations. In
contrast, in case $(ii)$, when $|q - 2q_0| \ll {\cal O}(1/a)$, the
corresponding terms proportional to $\lambda_D$ become slowly
varying and contribute to the dynamics. These terms emulate the
presence of a transverse effective field, causing electrons to flip
their spins along the direction of the combined uniform Rashba and
Dresselhaus fields. Turning to case $(iii)$, with $|q - 2k_{F}|  \ll
{\cal O}(1/a)$ but with $|q \pm 2q_0|  \simeq  {\cal O}(1/a)$, one
now finds that the terms proportional to $\lambda_D$ are washed away
upon integration, while the terms proportional to $\lambda_R$ or to
$\mu_{\text{mod}}$ survive. This implies that backscattering and CDW
correlations come into play, dramatically changing the physics: A
band gap opens at all four Fermi points, causing a {\em transition
to a nonmagnetic insulating state.} Finally, in case, $(iv)$, {\em
all} terms in Eq. (16) contribute to the integrated Hamiltonian,
leading to a rather complex theory. This case, however, where $k_F$
and $q_0$ both approach $q$, requires a fine tuning of both the
electron density (upon which $k_F$ depends) {\em and} the uniform
Rashba interaction in Eq. (\ref{Rashba2D}) (upon which $q_0$
depends). This case is expected to be hard to realize in an
experiment, and in the following we shall focus on the more
accessible case $(iii)$.

\subsection{Bosonization picture: Band insulator from modulated Rashba interaction}

To see how the spectacular effect driven by the modulated Rashba
interaction comes about (case $(iii)$ in the previous section), it
is useful to bosonize the theory. Using standard bosonization, we
write the right- and left-moving fermionic fields as
\bea R^{\phantom{\dagger}}_{\tau}(x)&=&\frac{\eta_{\tau}}{\sqrt{2\pi
a}} e^{{\it i}\sqrt{\pi}[\varphi_{\tau}(x)+\vartheta_{\tau}(x)]}, \label{bos1}\\
L^{\phantom{\dagger}}_{\tau}(x)&=&\frac{\bar{\eta}_{\tau}}{\sqrt{2\pi
a}} e^{{\it i}\sqrt{\pi}[\varphi_{\tau}(x)-\vartheta_{\tau}(x)]},
\label{bos2} \eea
where $\varphi_{\tau}(x)$ and $\vartheta_{\tau}(x)$ are dual bosonic
fields satisfying $\partial_t \varphi_{\tau} = v_{F} \partial_x
\vartheta_{\tau}$, and where $\eta_{\tau}$ and $\bar{\eta}_{\tau}$
are Klein factors which keep track of the fermion statistics for
electrons in different branches \cite{Giamarchi_book_04}.

Inserting the bosonized forms of $R^{\phantom{\dagger}}_{\tau}(x)$
and $L^{\phantom{\dagger}}_{\tau}(x)$ into Eq.
(\ref{linear-Hamiltonian}) and carrying out some simple algebra, one
obtains the bosonized Hamiltonian
\begin{multline} H_0^{\prime} + H_{DR}^{\prime}\!+\!H_\text{mod}^{\prime} = \sum_{\tau}\int
dx\,\Big\{\,\frac{v_{F}}{2}[\,(\partial_{x}\vartheta_{\tau})^{2}+(\partial_{x}\varphi_{\tau})^{2}\,] \\
+ \sum_{j=\pm1}\Big[ \frac{\lambda_{R}}{\pi
a}\sin \left((q+2jk_{F})x+\pi\nu+\sqrt{4\pi}\varphi_{\tau}\, \right) \\
- \frac{\mu_{\text{mod}}}{2\pi a}\sin
\left((q+2jk_{F})x+\sqrt{4\pi}\varphi_{\tau} \right) \Big] \,\Big\}.
\label{BosHam}
\end{multline}

It is useful to cast the Hamiltonian in Eq. (\ref{BosHam}) on the
more compact form
\begin{multline} \label{compact}
H_0^{\prime} + H_{DR}^{\prime}\!+\!H_\text{mod}^{\prime} \!=\!
\sum_{\tau}
\int \!dx\,\Big\{\,\frac{v_{F}}{2}[\,(\partial_{x}\vartheta_{\tau})^{2}\!+\!(\partial_{x}\varphi_{\tau})^{2}\,]  \\
\!+\!  \frac{M_{R}}{\pi a}
\!\sum_{j=\pm1}\!\cos[(q+2jk_{F})x\!+\!\phi_{0}\!+\!\sqrt{4\pi}\varphi_{\tau}]\Big\},
\end{multline}
where
\bea M_{R}&=&\sqrt{\lambda_{R}^{2}+
\mu_{\text{mod}}\lambda_{R}\cos(\pi\nu)+\mu_{\text{mod}}^{2}/4},
\label{MR}\\
\phi_{0}&=&-\arctan\left(\frac{\mu_{\text{mod}}+2\lambda_{R}\cos(\pi\nu)}{2\lambda_{R}\sin(\pi\nu)}\right).
\label{phi0} \eea

For the case that we are interested in, i.e. with $|q-2k_{F}| \ll
{\cal O}(1/a)$, the $j=-1$ component of the modulated term in Eq.
(\ref{compact}) comes into play \cite{FootnoteOnComponent}. For this
case we can gauge out the small term $\propto x$ from the argument
of the cosine by the transformation $(q-2k_{F})x +\phi_0 +
\sqrt{4\pi}\varphi_{\tau} \rightarrow \sqrt{4\pi}\varphi_{\tau}$ and
rewrite the Hamiltonian as $H_0^{\prime} +
H_{DR}^{\prime}\!+\!H_\text{mod}^{\prime}=\sum_{\tau}\int dx\, {\cal
H}_{\text{bos},\tau}(x)$, with Hamiltonian densities
\bea {\cal H}_{\text{bos},\tau} &=&
\frac{v_{F}}{2}[\,(\partial_{x}\varphi_{\tau})^{2}+(\partial_{x}\vartheta_{\tau})^{2}\,]-
\frac{\mu_{{\it eff}}}{\sqrt{\pi}}\cdot\partial_{x}\varphi_{\tau}
\nonumber\\
&+&\frac{M_{R}}{\pi
a}\cos(\sqrt{4\pi}\varphi_{\tau}),\label{H-tau_BOS} \eea
where
\begin{equation}
\mu_{\mathrm{eff}}=v_{F}(2k_{F}-q)/2 \label{mueff}
\end{equation}
serves as an effective chemical potential.
By tuning the density of electrons so that $\mu_{\mathrm{eff}}=0$,
the system is seen to be governed by two commuting sine-Gordon
models \cite{Coleman} with interaction terms $\cos(\beta \varphi_+)$
and $\cos(\beta \varphi_-)$ respectively, where $\beta^{2}=4\pi$. As
follows from the exact solution of the sine-Gordon model \cite{DHN},
in this case the excitation spectrum is gapped and consists of {\em
solitons} and {\em antisolitons} with masses $M_{+}=M_{-}=M_{R}$
(together with soliton-antisoliton bound states, so called {\em
breathers}, with masses $\ge M_R$). A soliton (or antisoliton)
corresponds to a configuration of the field $\varphi_{\tau}$, for a
given component $\tau$, that connects two neighboring minima
$\sqrt{4\pi} \varphi^{0}_{\tau}= \pi + 2\pi n\, (n \in {\cal Z})$ of
the functional potential $V[\varphi_{\tau}]=
M_{R}\cos(\sqrt{4\pi}\varphi_{\tau})$. The previous field
configurations define the set of possible ground states of
$\varphi_{\tau}$ with vacuum expectation values $\langle
\varphi^{0}_{\tau}\rangle = \sqrt{\pi}(1/2+n)$. For example, a field
configuration where $\varphi_{\tau}(-\infty) = \sqrt{\pi}/2 \
[3\sqrt{\pi}/2]$ and $\varphi_{\tau}(\infty) = 3\sqrt{\pi}/2 \
[\sqrt{\pi}/2]$ supports a soliton [antisoliton] with fermion number
$N_{\tau} = 1 \, [-1]$, defined by \be N_{\tau} =
\frac{1}{\sqrt{\pi}} \int_{-\infty}^{\infty} dx \, \p_x
\varphi_{\tau} (x). \label{fermi-number} \ee
The charge and spin quantum numbers of the single-particle
excitation are given by
\be Q = N_{+} + N_{-},\qquad S_z = \frac{1}{2}(N_{+} - N_{-}). \ee
The simplest single-particle excitation is obtained by considering a
soliton or antisoliton in the spin $\tau = +$ component, keeping the
ground state unperturbed for the spin $\tau = -$ component: $N_{+} =
\pm 1$, $N_{-} = 0$. Such an excitation has charge $Q= \pm 1$ and
spin $S_z=\pm 1/2$ (with spin projections $\tau = \pm$ along the
direction of the momentum-dependent combined uniform Rashba and
Dresselhaus fields). Thus, {\it the elementary excitations of the
system are free massive fermions with mass $M_R$, each carrying unit
charge and spin 1/2.} It follows that the joint action of the
modulated Rashba coupling and the chemical potential, with the
electron density tuned so as to satisfy the commensurability
condition $\mu_{\mathrm{eff}} = 0$, turns the electron gas into an
effective band insulator. The corresponding band gap is equal to the
doubled mass of the single-particle excitation, $\Delta = 2M_R$,
since conservation of charge and spin requires the simultaneous
excitation of a soliton and an anti-soliton. Note from Eq.
(\ref{MR}) and the definition of $\lambda_R$ after Eq.
(\ref{linear-Hamiltonian}) that the effect of the Dresselhaus
interaction is to reduce the gap. Fortunately, as we shall show in
Sec. V, this unwanted effect (from the point of view of spintronics
applications) is negligible when compared to the stronger Rashba
interaction.

\subsection{Bosonization picture in the spin-charge basis}

The nature of the metal-insulator transition becomes more
transparent if we treat the model in a basis with charge $(c)$ and
spin $(s)$ bosons $-$ the standard basis in which to include effects
of electron-electron interactions \cite{Giamarchi_book_04}. Thus
introducing the dual {\em charge fields}
\begin{equation}
\varphi_{c} = {\textstyle \frac{1}{\sqrt{2}}} (\varphi_{+} +
\varphi_{-}), \qquad \vartheta_{c} = {\textstyle \frac{1}{\sqrt{2}}}
(\vartheta_{+} + \vartheta_{-}) \label{bos_charge}
\end{equation}
and {\em spin fields}
\begin{equation}
\varphi_{s} = {\textstyle \frac{1}{\sqrt{2}}} (\varphi_{+} -
\varphi_{-}),\qquad \vartheta_{s} = {\textstyle \frac{1}{\sqrt{2}}}
(\vartheta_{+} - \vartheta_{-}), \label{bos_spin}
\end{equation}
 some simple
algebra on Eq. (\ref{H-tau_BOS}) yields that $H_0^{\prime} +
H_{DR}^{\prime}\!+\!H_\text{mod}^{\prime}= \int dx\,[{\cal H}_{0c} +
{\cal H}_{0s} + {\cal H}_{cs}]$, with
\bea {\cal H}_{0c} & = & \frac{v_{F}}{2}
[(\partial_{x}\varphi_{c})^2 +(\partial_x
\vartheta_{c})^2]-\sqrt{\frac{2}{\pi}}\mu_{{\it
eff}}\partial_{x}\varphi_{c}\, ,\label{SGc}\\
{\cal H}_{0s} &=&   \frac{v_{F}}{2} [(\partial_{x}\varphi_{s})^2
+(\partial_x \vartheta_{s})^2\big]\, ,\label{SGs} \\
{\cal H}_{cs}  &=&  \frac{2M_{R}}{\pi
a}\cos(\sqrt{2\pi}\varphi_{c})\cos(\sqrt{2\pi}\varphi_{s})\, .
\label{H_CS_BOS} \eea

At $\mu_{\text{eff}}=0$ Eqs. (\ref{SGc})-(\ref{H_CS_BOS}) describe
two bosonic charge and spin fields coupled by the strongly
(renormalization-group) relevant operator ${\cal H}_{cs}$ (of
scaling dimension 1). This operator drives the system to a
strong-coupling regime where the charge and spin fields are pinned
at their ground state expectation values. Therefore, at
$\mu_{\text{eff}}=0$, both charge and spin excitations develop a gap
(let us call them $M_c$ and $M_s$, respectively) and the system
becomes a nonmagnetic insulator, consistent with the finding in the
previous section where the system develops a gap also in the bosonic
$\tau=\pm$ basis. As $\mu_{\text{eff}}$ is tuned away from zero the
influence of the operator ${\cal H}_{cs}$ in Eq. (\ref{H_CS_BOS})
gets weaker and eventually averages to zero upon integration when
$\mu_{\text{eff}}$ exceeds the insulator band gap given by the mass
$M_c$ of charge excitations. At this point, the system then turns
metallic. Therefore, the competition between the chemical potential
term and the commensurability energy drives a continuous
insulator-to-metal transition from a gapped phase at
$\mu_{\mathrm{eff}} < \mu_{\mathrm{eff}}^{c}$ to a gapless phase at
$\mu_{\mathrm{eff}} > \mu_{\mathrm{eff}}^{c} =
M_c$\cite{C_IC_transition}. The (de-)tuning of $\mu_{\text{eff}}$ in
our Eq. (\ref{mueff}) can be achieved by changing either $k_F$ or
$q$, i.e. the band filling $\nu$ ($k_F=\pi\nu/a$) or the wave length
$\lambda$ of the gate modulation ($q =2\pi/\lambda$). Either
alternative poses its own experimental difficulties, although we
expect that the band filling is more easily controllable, using a
back gate with a variable voltage. Therefore, we shall hereafter
assume that the tuning mechanism is provided by an adjustable band
filling. Thus $-$ rephrasing this in a language closer to experiment
$-$ by detuning the voltage of the backgate of the device so that
the electron density $n_s$ of the quantum well is shifted from the
value $\pi/2\lambda^2$, one will observe a transition from a
nonmagnetic  insulating state into a metallic
phase\cite{densityfootnote}. In this phase the electrons in the wire
exhibit ordinary Fermi liquid behavior with gapless quasiparticle
excitations. This kind of transition belongs to the universality
class of commensurate-to-incommensurate transitions
\cite{Giamarchi_book_04}: The conductivity $\sigma$ close to the
transition scales as $\sigma \sim (\mu-\mu_c)^{1/2}$, with the
compressibility $\kappa$ diverging as $\kappa \sim
(\mu-\mu_c)^{-1/2}$, before dropping to zero on the insulating side.

The insulator-to-metal transition just discussed corresponds to the
picture put forward by Schulz in Ref. \onlinecite{Schulz} where a
Hamiltonian similar to that defined by our Eqs. (\ref{SGc}) -
(\ref{H_CS_BOS}) is refermionized into a two-band model (cf. Eq. (4)
in Ref. \onlinecite{Schulz}). The two bands are separated by a gap
$\Delta$, with a chemical potential $\mu_0=0$ corresponding to a
completely filled lower band. In other words, in this state the
system is a band insulator with a gap $\Delta$. For $\mu_0$ smaller
than the critical value $-\Delta/2$, holes are introduced at the top
of the lower band, whereas for $\mu_0$ larger than $\Delta/2$,
electrons are added to the bottom of the upper band; in both cases
the system becomes metallic. This refermionized picture thus makes
it clear that it takes a {\em finite} critical $\mu_0$ for the
transition to occur: by tuning the chemical potential to zero the
system not only develops a gap but also a rigidity that sustains the
gap when the system is shifted away from commensurability.

Going back to Eqs. (\ref{SGc}) - (\ref{H_CS_BOS}), it is instructive
to see how the single-fermion excitations obtained in the previous
section can be reconstructed in the present spin-charge basis.  Here
we follow a route developed in Ref. \onlinecite{FGN} in studies of a
similar problem in the case of the ionic Hubbard model. First note
that only the relative sign between $\langle \cos(\sqrt{2\pi}
\varphi_c)\rangle$ and $\langle \cos(\sqrt{2\pi} \varphi_s)\rangle$
in Eq. (\ref{H_CS_BOS})  is fixed to be negative (so as to minimize
$\langle {\cal H}_{cs} \rangle$). Thus, there are two possibilities
for the ground state charge and spin field expectation values:
\bea \mbox{I}: \  \langle \varphi_c \rangle & \! = \! & \sqrt{2\pi}m
\ \mbox{and}
\ \langle \varphi_s \rangle =  \sqrt{\frac{\pi}{2}}(2n+1), \label{sets1}  \\
\mbox{II}: \  \langle \varphi_c \rangle & \! = \! &
\sqrt{\frac{\pi}{2}}(2m +1) \ \mbox{and} \ \langle \varphi_s \rangle
=  \sqrt{2\pi}n, \label{sets2} \eea
with $m, n \in {\cal Z}$. To obtain the single-fermion excitations
one has to consider field configurations that connect two
groundstates that belong to {\em distinct} sets I  (Eq.
(\ref{sets1})) and II (Eq. (\ref{sets2})). As an example, a field
configuration that connects $\varphi_c =0$ with $\varphi_c =
\sqrt{\pi/2}$ in the charge sector and $\varphi_s=\sqrt{\pi/2}$ with
$\varphi_s = \sqrt{2\pi}$ in the spin sector corresponds to an
excitation with charge and spin quantum numbers \cite{GNT}
\bea
Q & = & \sqrt{\frac{2}{\pi}} \int_{-\infty}^{\infty} dx\, \partial_x \varphi_c(x) = 1, \label{fermionQ}\\
S_z & = & \frac{1}{\sqrt{2\pi}} \int_{-\infty}^{\infty} dx\,
\partial_x \varphi_s(x) = 1/2, \label{fermionS} \eea
i.e. a massive fermion (of mass $M_{R}$), which is the elementary
excitation in the band insulator. To obtain a pure charge or spin
excitation one must consider field configurations that connect
groundstates {\em within} the sets I and II. For example, given set
I in Eq. (\ref{sets1}), we can lock the charge at $\varphi_c=0$ and
consider a {\em spin soliton} connecting the groundstates at
$\varphi_s = \sqrt{\pi/2}$ and $\varphi_s = 3\sqrt{\pi/2}$. Such an
excitation carries charge $Q=0$ and spin
\begin{equation}  \label{spinsoliton}
S_z = \frac{1}{\sqrt{2\pi}} \int_{-\infty}^{\infty} dx\, \partial_x
\varphi_s(x) = 1.
\end{equation}
In the noninteracting case considered here it is clear that that
this excitation is built from two massive fermions with opposite
charge and the same spin. Similarly, a {\em charge soliton} can be
obtained by locking the spin at one of the possible groundstates and
consider a charge field configuration that connects, say, $\varphi_c
= 0$ and $\varphi_c = \sqrt{2\pi}$. This excitation carries charge
\begin{equation} \label{chargesoliton}
Q = \sqrt{\frac{2}{\pi}} \int_{-\infty}^{\infty} dx\, \partial_x
\varphi_c(x) = 2
\end{equation}
while having zero spin, being built from two massive fermions with
the same charge and opposite spin.

Following this logic, a derivation of $M_c$ and $M_s$ should give
$M_c=M_s=2M_R$. As we shall see, the mean-field approach used in the
next section to evaluate $M_c$ and $M_s$ gives a slightly
overestimated value. We shall return to this issue below, and show
how it can be resolved by a proper regularization procedure.

The opening of a gap for both charge and spin excitations \cite{SSH}
at commensurability, $\mu_{\text{eff}} = 0$, reflects the fact that
the system has turned into a {\em nonmagnetic band insulator}. Using
the standard bosonized expression for the charge density, \cite{GNT}
\begin{multline} \rho_{c}(x) \simeq
\frac{1}{\sqrt{2\pi}}\partial_{x}\varphi_{c} \\ +
A\sin(\sqrt{2\pi}\varphi_{c}+2k_{F}x) \cos(\sqrt{2\pi}\varphi_{s})
 \label{upDown_density} \end{multline}
with $A$ a constant, one verifies from Eqs. (\ref{sets1}) and
(\ref{sets2}) (which apply at $\mu_{\text{eff}}=0$), that the ground
state of the system corresponds to a {\em CDW-type band insulator},
with long-range charge-density modulation
\bea \rho_{c}(x) \simeq \rho^{m}_{c}\sin(2k_{F}x), &
 \label{upDown_density} \eea
where
 \bea &\rho^{m}_{c} \sim \la\cos(\sqrt{2\pi}\varphi_{c})\ra\la\cos(\sqrt{2\pi}\varphi_{s})\ra\, .
&
 \label{upDown_density} \eea
As should be clear from the non-conservation of spin in the presence
of the spin-orbit interactions, the massiveness of the spin
excitations does not correspond to the formation of a spin density
wave (SDW). Indeed, by writing down the bosonized
expression\cite{GNT} for a SDW with spin projection along the
direction of the combined uniform Rashba and Dresselhaus fields,
$\rho_s(x) \simeq (1/\sqrt{2\pi}) \partial_x\varphi_s +
B\cos(\sqrt{2\pi} \varphi_c + 2k_Fx)\sin(\sqrt{2\pi} \varphi_s)$,
with $B$ a constant, one immediately verifies from Eqs.
(\ref{sets1}) and (\ref{sets2}) that it has no amplitude for a
long-range $2k_F$ modulation.

\subsection{Bosonic mean-field theory in \\ the spin-charge basis}

To pave the ground for including electron-electron interactions into
the problem, we next decouple the interaction term ${\cal H}_{cs}$
in Eq. (\ref{H_CS_BOS}) in a mean-field manner by introducing
\begin{eqnarray}\label{Mc}
\label{m_{c}}
m_{c}&=& 2M_{R} | \langle \cos(\sqrt{2\pi}\varphi_{s})\rangle |\,\, ,\\
\label{m_{s}} m_{s}&=& 2M_{R} |\langle
\cos(\sqrt{2\pi}\varphi_{c})\rangle |\, . \label{Ms}
\end{eqnarray}
Note that the mean-field decoupling is well controlled since, at the
strong-coupling fixed point,  fluctuations are strongly suppressed
by the pinning of the charge and spin bosons. Using Eqs. (\ref{Mc})
and (\ref{Ms}), we find that the mean-field version of the bosonized
Hamiltonian $H_0^{\prime} +
H_{DR}^{\prime}\!+\!H_\text{mod}^{\prime}= \int dx\,[{\cal H}_{0c} +
{\cal H}_{0s} + {\cal H}_{cs}]$, defined in Eqs. (\ref{SGc}) -
(\ref{H_CS_BOS}), can be written as $H_{\text{mean}} = \int \! dx \,
[{\cal H}_{c} + {\cal H}_{s}]$ with
\begin{eqnarray}
{\cal H}_{c}&=&{v_{F} \over 2}[(\partial_x
\varphi_{c})^2+(\partial_{x}\vartheta_{c})^2]+\frac{m_{c}}{\pi
a}\cos(\sqrt{2\pi} \varphi_{c})-\nonumber\\
&& -\sqrt{\frac{2}{\pi}}\mu_{\reff}\partial_{x}\varphi_{c},
\label{Free_Bos_MF_Charge}\\\nonumber\\
{\cal H}_{s}&=&{v_{F} \over 2}[(\partial_x
\varphi_{s})^2+(\partial_{x}\vartheta_{s})^2]+\frac{m_{s}}{\pi
a}\cos(\sqrt{2\pi} \varphi_{s})\, . \label{Free_Bos_MF_Spin}
\end{eqnarray}

When $\mu_{\reff} = 0$, the Hamiltonian defined by Eqs.
(\ref{Free_Bos_MF_Charge}) and  (\ref{Free_Bos_MF_Spin}) is again
given by a sum of two decoupled sine-Gordon models (cf. Eq.
(\ref{H-tau_BOS})). However, the dimensionalities of the
$\cos(\beta\varphi)$ operators at $\beta^{2}=2\pi$ [spin-charge
basis, Eqs. (\ref{Free_Bos_MF_Charge}), (\ref{Free_Bos_MF_Spin})]
and $\beta^{2}=4\pi$ [$\tau = \pm$ basis, Eq. (\ref{H-tau_BOS})] are
different.

By exploiting  the exact solution of the sine-Gordon model, we can
easily estimate the size of the insulating gap in the spin-charge
basis \cite{JHLM-H_2007}. The excitation spectra of Eq.
(\ref{Free_Bos_MF_Charge}) at $\mu_{\reff} = 0$ and Eq.
(\ref{Free_Bos_MF_Spin}) consist of solitons and antisolitons with
masses $M_c$ and $M_s$, respectively (in addition to the charge and
spin breathers with masses bounded below by $M_c$ and $M_s$,
respectively). These charge and spin soliton masses are related to
the ``bare" masses $m_c$ and $m_s$ in Eqs.
(\ref{Free_Bos_MF_Charge}) - (\ref{Free_Bos_MF_Spin}) by
\cite{Al_B_Zamolodchikov_95}
\be\label{Al-B-Z_1} M_{\kappa}/\Lambda = {\cal
C}_{0}\left(m_{\kappa}/\Lambda\right)^{2/3},  \ \kappa = c, s. \ee
with $\Lambda$ an energy cutoff that blocks excitations into the
second conduction channel of the quantum wire (for details, see Sec.
V).

The ground state expectation values of
$\cos(\sqrt{2\pi}\varphi_{\kappa})$ are in turn given by
\cite{Luk_Zam_97}
\be\label{A-B-Z_1} | \la \, \cos \sqrt{2\pi} \varphi_{\kappa} \, \ra
| = {\cal C}_{1}\left(M_{\kappa}/\Lambda\right)^{1/2}, \ \ \kappa =
c,s, \ee
with ${\cal C}_{0}\approx 1.4$ and ${\cal C}_{1}\approx 1.0.$ (For
details, see Appendix A.) By combining Eqs. (\ref{Al-B-Z_1}) and
(\ref{A-B-Z_1}) with (\ref{Mc}) and (\ref{Ms}) one reads off that
\be\label{Delta_c-Delta_s} M_{c}=M_{s} = 2\gamma M_{R}, \ee
with $\gamma = {\cal C}^{3/2}_{0}{\cal C}_{1}\simeq 1.7$.

Note that charge and spin solitons, though formally decoupled, move
with the same velocity $v_{F}$ (cf. Eqs. (\ref{Free_Bos_MF_Charge})
- (\ref{Free_Bos_MF_Spin})), a record of their composite nature
since, as demonstrated in Sec. III.C, charge and spin excitations
are built from single fermions of mass $M_{R}$, unit charge and spin
$S=1/2$, the latter being the elementary excitations in the
$\tau=\pm$ basis of Sec. III.B. Thus, the mean-field treatment in
the spin-charge basis faithfully captures the character of the
excitations. However, the size of the two-particle gap, $M_c$ or
$M_s$ as given in Eq. (\ref{Delta_c-Delta_s}), gets overestimated by
a factor of 1.7 when compared to the result $\Delta = 2M_R$,
obtained in Sec. III.B. As we shall show in the next section, the
factor of 1.7 can be removed by introducing a regularized form of
the gap.

Recall that tuning the effective chemical potential away from zero
``closes" the band gap, and thus drives an insulator-to-metal
transition by depinning the charge field from its ground state
expectation value. The combined Eqs. \!(\ref{Ms}),
\!(\ref{Al-B-Z_1}), and (\ref{A-B-Z_1}) reveal that, in the process,
the spin sector becomes gapless as well, as $M_{s} \sim \langle \cos
(\sqrt{2\pi}\varphi_{c})\rangle = 0$.

\subsection{Functional behavior of the effective band gap}

Having established that the masses of the charge- and spin
excitations in the non-interacting theory, $M_c$ and $M_s$
respectively, are determined by the single-fermion mass $M_R$, let
us return to Eq. (\ref{MR}) to analyze its dependence on the
relative phase between the two modulations and their amplitudes
$\lambda_R$ and $\mu_{\text{mod}}$. As emphasized in the previous
sections, the mass $M_R$ is a key parameter of our theory, encoding
the effective band gap $\Delta = 2M_R$ in the insulating state of
noninteracting electrons.

In Sec. V, when analyzing a generic gated heterostructure, we shall
see that both $\lambda_R$ and $\mu_{\text{mod}}$ depend linearly on
the voltage $V_{\text{G}}$ of the top gates (cf. Fig. 1). We may
thus write $\lambda_R=c_1V_{\text{G}}$ and
$\mu_{\text{mod}}=c_2V_{\text{G}}$, where $c_1$ and $c_2$ are
constants depending on the details of the setup and of the sample.

To analyze the gap behavior it is important to distinguish the two
ways in which the parameters $\lambda_R$ and $\mu_{\text{mod}}$ can
be varied: One possibility is to consider (i) a {\em fixed system}
(i.e. keeping $c_1$ and $c_2$ fixed) and varying the gate voltage
$V_{\text{G}}$; alternatively, one may consider (ii) {\em different
systems} but keeping the gate voltage fixed, e.g. by testing
different samples from an ensemble of properly gated
heterostructures (all of which satisfy the commensurability
condition $|q-2k_{F}| \ll {\cal O}(1/a)$).

Let us start by investigating the possibility (i). In this case, we
can rewrite $\mu_{\text{mod}}=(c_2/c_1)\lambda_R$ and Eq. (\ref{MR})
as
\begin{equation}
M_R=c(\nu)|\lambda_R| \label{MR2}
\end{equation}
where $c(\nu)=\sqrt{1+(c_2/c_1)\cos(\pi\nu)+(c_2/c_1)^2/4}$ is a
system specific parameter adjustable by the band filling $\nu$
(which, in turn, can be varied by a back gate with a variable
voltage).

\begin{figure}[htpb]
\begin{center}
\includegraphics[scale=0.25]{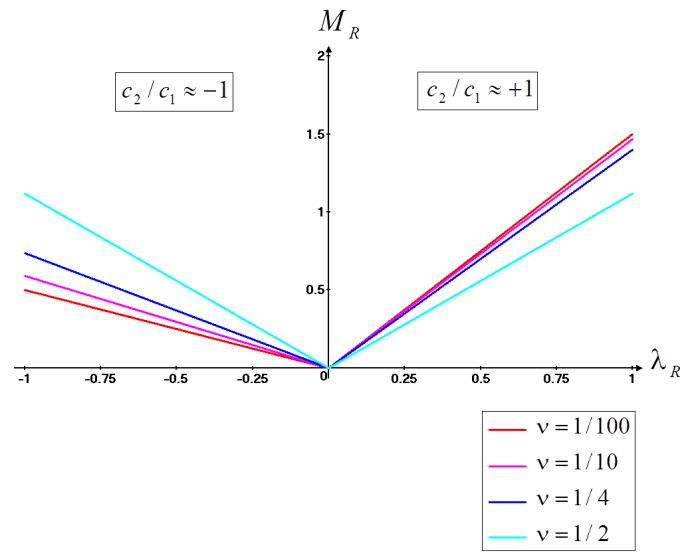}
\caption{(Color online) $M_{R}$ as a function of $\lambda_R$ (common
arbitrary units) for given values of the band-filling $\nu$. ``In
phase" [``Out of phase"] modulations correspond to $c_2/c_1\simeq1$
[$c_2/c_1\simeq-1$]}. \label{Fig:fig2}
\end{center}
\end{figure}

Figure \ref{Fig:fig2} shows $M_{R}$ as a function of $\lambda_R$ for
band fillings $\nu = 1/100, 1/10, 1/4, 1/2$. The reason for
considering systems only up to half-filling is the following: Due to
the commensurability condition $|q-2k_{F}| \ll {\cal O}(1/a)$, the
values of the filling $\nu$ in Figure \ref{Fig:fig2} correspond to
modulation wave lengths $\lambda = 100a, 10a, 4a,$ and $2a$,
respectively (as seen from the relations $q=2\pi/\lambda$ and
$k_F=\pi\nu/a$). Since the ultrasmall gates that we propose to be
used for producing the modulation each has a spatial extension
$\lambda/2$ along the quantum wire (c.f. Figure 1), it follows that
$\nu = 1/2$ sets an upper (and in practice, unattainable) physical
limit for possible band fillings: If $\nu > 1/2$, the dimension of a
gate would have to become subatomic. In fact, as can be gleaned from
the experimental data cited in Sec. V, our theory would likely break
down already for band fillings around $\nu \approx 1/3$ since at
larger fillings higher subbands will come into play, causing subband
mixing. As we shall also see in Sec. V, the 1D band filling with
present-day semiconductor heterostructures is typically around $\nu
\approx 1/10$, implying a gate extension of a few nanometers.
Already this presents a challenge to the experimentalist.

The plots in Figure \ref{Fig:fig2} are shown for $\lambda_R$ running
from $-1$ to $+1$, thus accounting for the two possible phase
relations between the Rashba and the chemical potential modulations.
For ``in phase" [``out of phase"] modulations, the plots are shown
for test systems where $c_2/c_1\simeq1$ [$c_2/c_1\simeq-1$].

We see that for both ``in phase" and ``out of phase" modulations,
the gap is an increasing linear function of $|\lambda_{R}|$ with
slope depending on the band filling. The increase of the gap is
consistent with the phenomenological expectation that the insulating
state gets more stable as the pinning Rashba interaction goes
stronger, and is in agreement also with the corresponding result in
Ref. \onlinecite{JJF_Paper_09}. There, however, the modulation of
the charge density was not taken into account and, thus, the
formalism did not capture the gap dependence on the band filling.
The split of a single gap line for different values of $\nu$, as
manifest in Figure \ref{Fig:fig2}, is an interesting feature of the
system resulting from the combination of the modulated Rashba
interaction and CDW correlations.


Another interesting aspect of the gap behavior is that, given a
certain band filling $\nu$ and a value for
$\mid\!\!\lambda_R\!\!\mid$, the gap for ``in phase" modulations is
larger than for ``out of phase" modulations, implying a stronger
localization effect when the the Rashba interaction and the chemical
potential act in ``unison''. The difference
$M_R(\lambda_R)-M_R(-\lambda_R)$ goes to zero as $\nu$ approaches
1/2 (half-filled band).

Turning now to case (ii), we can define a new variable
$\delta\equiv\lambda_R/\mu_{\text{mod}}=c_1/c_2$ that characterizes
a particular setup, material, or design and is independent of the
value of the applied gate voltage. With that we can rewrite Eq.
(\ref{MR}) as
\begin{equation}
m_R=\sqrt{\delta^2+\cos(\pi\nu)\delta+1/4}\label{MR3}
\end{equation}
where $m_R=M_R/\mu_{\text{mod}}$.

\begin{figure}[htpb]
\begin{center}
\includegraphics[scale=0.25]{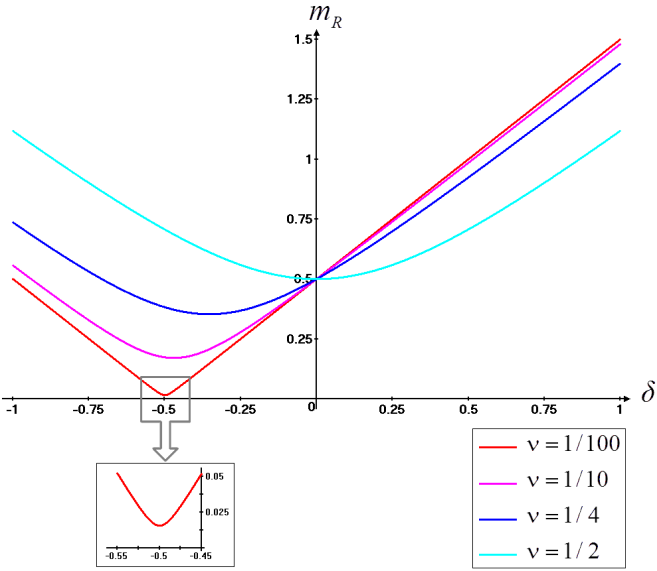}
\caption{(Color online) $m_{R}$ as a function of the ratio
$\delta\equiv\lambda_{R}/\mu_{\text{mod}}$ for given values of the
band-filling $\nu$.} \label{Fig:fig3}
\end{center}
\end{figure}

Figure \ref{Fig:fig3} shows $m_{R}$ as a function of $\delta$ for
the same band fillings $\nu = 1/100, 1/10, 1/4, 1/2$ used in case
(i). The plots are shown for $\delta$ running from $-1$ to $+1$,
accounting for ``in phase" and ``out of phase" Rashba and chemical
potential modulations.

To understand the behavior revealed by Figure \ref{Fig:fig3}, let us
first look at the right-half of the graph where the Rashba and
chemical potential modulations are ``in phase". The plots here show
a monotonic increase of $m_R$ with $\delta$, implying that with
devices where the top gate voltages $V_{\text{G}}$ have been tuned
so as to produce the same fixed value of $\mu_{\text{mod}}$, the gap
will be larger in the device with the larger value of $\lambda_R$.
This behavior is equivalent to that obtained in case (i).

Now turn to the left-half of the graph where the Rashba and chemical
potential modulations are ``out of phase". Again comparing devices
having the same band filling and for which the magnitude
$V_{\text{G}}$ of the gate voltages have been tuned to give the same
$\mu_{\text{mod}}$, we observe an unexpected feature. Let us follow
what happens when going through different devices by moving along a
curve with a fixed band filling (smaller than 1/2): the gap first
decreases with the strength of the Rashba interaction until it
reaches a minimum at $\delta^{\star}=\cos(\pi\nu)/2$; past this
value the ``normal" increasing behavior is recovered. This
\emph{crossover} behavior gets more pronounced for smaller values of
$\nu$, i.e. as the system goes more diluted. For $\nu=1/100$, for
example, the crossover almost annihilates the gap at
$\delta\simeq0.5$ (but not completely as can be seen by zooming in
around that point).

To understand how this phenomenon comes about, consider a {\em
Gedanken} experiment with a single device where $\lambda_R$ is
allowed to vary while $\mu_{\text{mod}}$ is kept fixed. In the case
with anti-phased modulations, the chemical potential [Rashba
potential] will have a maximum [minimum] in the middle of the
segment, call it ``A", below one of the small positively charged
gates (and vice versa for a neighboring gate-free segment, call it
``$B$" (cf. Fig. 1)). Consider first the case with $\lambda_R=0$.
Here all electrons will reside in the $A$-regions since this
configuration is energetically more advantageous. This causes a
localization of electrons, with a gap $M_R = \mu_{\text{mod}}/2$ as
seen in Figure \ref{Fig:fig3} for $\delta=0$. Now turn on
$\lambda_R$. To take advantage of the spin-flip Rashba hopping some
electrons will start migrating into the $B$-regions. This weakens
the localization effect of the chemical potential modulation, with
the result that the gap {\em decreases}. By successively increasing
$\lambda_R$, more and more electrons will migrate into the
$B$-regions, and for a sufficiently large value of $\lambda_R$,
equal to $\mu_{\text{mod}}\cos(\pi\nu)/2$, {\em all} electrons will
reside in the $B$-regions. From that point on the gap will increase
monotonically as $\lambda_R$ (or $\gamma_R$) is further increased,
as displayed in Figure \ref{Fig:fig3}. For fillings close to the
upper physical limit, that is 1/2, a small $\lambda_R$ is enough for
a complete electronic migration between the chemical potential
$A$-regions to the spin-orbit coupling $B$-regions while, for a more
dilute system, the process is ``slower", demanding a larger
$\lambda_R$. The two extreme cases are $\nu=1/2$ for which the
$A$-regions are emptied right away for any nonzero $\lambda_R$ and
$\nu\rightarrow0$ for which the necessary $\lambda_R$ is (still
just) half the amplitude $\mu_{\text{mod}}$ of the original pinning
chemical potential.

In the next section we shall investigate how the electron-electron
interaction influences the results thus obtained.

\section{Interacting electrons}
\subsection{Adding interactions: Bosonization picture \\ in the spin-charge basis}

To incorporate the electron-electron interaction
$H_{\text{e-e}}^{\prime}$ in Eq. (\ref{e-eTransformed}) into the
bosonic theory we perform the same steps as in Sec. III.B, first
linearizing the spectrum around the four Fermi points and taking a
continuum limit. This yields $H_{\text{e-e}}^{\prime} = \int \, dx\,
{\cal H}_{\text{e-e}}(x)$, with
\begin{eqnarray}
{\cal H}_{\text{e-e}} &=& g_1^{\tau
\tau'}\!:\!R^{\dag}_{\tau}L^{\phantom{\dagger}}_{\tau}L^{\dag}_{\tau'}
R^{\phantom{\dagger}}_{\tau'}\!: +  \, {g}_{2}^{\tau \tau'}
\!:\!R^{\dag}_{\tau}R^{\phantom{\dagger}}_{\tau}
L^{\dag}_{\tau'}L^{\phantom{\dagger}}_{\tau'}\!: \nonumber  \\
&+&  \frac{g_{4}^{\tau \tau'}}{2}(
:\!L^{\dag}_{\tau}L^{\phantom{\dagger}}_{\tau}
L^{\dag}_{\tau'}L^{\phantom{\dagger}}_{\tau'}\!: + \, R
\leftrightarrow L), \label{LCHee}
 \end{eqnarray}
with $\tau, \tau' = \pm$ summed over, and where $g^{\tau\tau'}_{1}$
and $g^{\tau\tau'}_{2}$ are the amplitudes, respectively, for back
and forward (``dispersive'') scattering between electrons of
different chiralities, and $g^{\tau\tau'}_{4}$ is the amplitude for
forward scattering between electrons of equal chirality
\cite{Giamarchi_book_04}. Whereas the $g_2^{\tau \tau'}$ and
$g_4^{\tau \tau'}$ processes correspond to scattering with small
momentum transfer, the $g_1^{\tau \tau'}$ process transfers momentum
$k \sim 2k_F$. For a screened Coulomb interaction with a nonzero
screening length, the $g_1^{\tau \tau'}$ amplitude is therefore
quite small, and can usually be neglected. This is certainly so in
the present case since in a semiconductor structure the Coulomb
interaction is much smaller than the band width. It follows that in
this limit the $k \sim 2k_F$ scattering becomes marginally
irrelevant and renormalizes to zero at low energies. Importantly,
this conclusion is not invalidated by the presence of the spin-orbit
couplings \cite{GJPB_05}.  From now on we shall therefore consider
the simpler theory where the back scattering has been renormalized
away, i.e. with  $g^{\tau\tau'}_{1} \simeq 0$.

For a system at commensurate band-filling $\nu=1/2n$, with $n\ge 1$
an integer, the Hamiltonian density in Eq. (\ref{LCHee}) should be
supplemented by an umklapp term which describes the transfer of $2n$
electrons of equal chirality to the opposite Fermi point through
exchange of momentum with the lattice. As is well known, these
processes drive a transition to an insulating state at a critical
value of the Coulomb interaction determined by the number $2n$ of
electrons participating in the process \cite{GiamarchiMillis}.
However, the screened Coulomb interaction in a gated semiconductor
structure is too weak to support such a transition except at a
half-filled $(n=1)$ or possibly a quarter-filled $(n=2)$ band
\cite{Hausler}. This should be contrasted with the commensurability
condition $q = 2k_F$ for driving a metal-to-insulator transition via
a modulated Rashba interaction, as derived in Sec. III.B. Since
$q=2\pi/\lambda$, with $\lambda$ the wavelength of the Rashba
modulation, and $k_F=\nu \pi/a$, this condition translates to
$\lambda = 2na$ when $\nu=1/2n$. Thus, with a Rashba modulation
tuned to commensurability with $2k_F$, umklapp processes at $n=1,2$
could come into play only for a sequence of electrical gates of
near-atomic dimensions, $\lambda/2 \sim a$. For this reason we shall
neglect umklapp processes when studying the novel physics coming
from a  Rashba modulation.

Having disposed of backscattering and umklapp processes, the
remaining electron-electron interaction in Eq. (\ref{LCHee}) is now
easily bosonized using Eqs. (\ref{bos1}), (\ref{bos2}),
(\ref{bos_charge}), and (\ref{bos_spin}). The resulting expression
for the bosonized mean field theory representing the full $H'$ in
Eq. (\ref{transformed}) then takes the form
$H_{\text{mean}}^{\prime} = \int dx\,[\,{\cal H}_{c}^{\prime} +
{\cal H}_{s}^{\prime}\,]$, with
\begin{eqnarray}
{\cal H}_{c}^{\prime}&=&\int dx\Big\{{v_{c} \over
2}[(\partial_{x}\vartheta_{c})^2 +(\partial_x
\varphi_{c})^2]\nonumber\\
&&- \mu_{\reff}\sqrt{\frac{2K_c}{\pi}}\partial_{x}\varphi_{c}
+\frac{m_{c}}{\pi a}\cos(\sqrt{2\pi K_c} \varphi_{c})\Big\},
\label{Int_Bos_MF_Charge} \\
{\cal H}_{s}^{\prime}&=&\int dx\Big\{{v_{s} \over
2}[(\partial_{x}\vartheta_{s})^2 + (\partial_x
\varphi_{s})^2]\nonumber\\
&&+\frac{m_{s}}{\pi a}\cos(\sqrt{2\pi K_s} \varphi_{s})\Big\}\, .
\label{Int_Bos_MF_Spin}
\end{eqnarray}
where we have performed the field transformations $\varphi_i
\rightarrow \sqrt{K_i} \varphi_i$ and $\vartheta_i \rightarrow
\vartheta_i/\sqrt{K_i}, \, i=c,s$. A comparison with Eqs.
(\ref{Free_Bos_MF_Charge}) and (\ref{Free_Bos_MF_Spin}) shows that
$H_{\text{mean}}^{\prime}$  has the same structure as the mean field
theory for noninteracting electrons and is given by two decoupled
sine-Gordon models when the commensurability condition
$\mu_{\text{eff}} =0$ is satisfied. The electron-electron
interaction is encoded by the new parameters $v_i$ and $K_i, \,
i=c,s$, as well as by the reparameterization of the bare masses
$m_c$ and $m_s$  due to the transformation $\varphi_i \rightarrow
\sqrt{K_i} \varphi_i$ (cf. Eqs. (\ref{Mc}), (\ref{Ms})). In the
weak-interaction limit considered here, $v_i$ and $K_i$ can be given
explicit representations in terms of the scattering amplitudes in
Eq. (\ref{LCHee}). Introducing the conventional ``g-ology" notation
\cite{Giamarchi_book_04} $g_{\parallel} \equiv g^{\tau\tau}$ for
parallel spins and $g_{\perp} \equiv g^{\tau\,-\tau}$ for opposite
spins, one has that
\begin{equation}
v_{i}=v_{F}[(1+y_{4i}/2)^{2}-(y_{i}/2)^{2}]^{1/2}, \label{vi}
\end{equation}
\begin{equation}
K_{i}=\left[\frac{1+y_{4i}/2+y_{i}/2}{1+y_{4i}/2-y_{i}/2}\right]^{1/2},
\label{Ki}
\end{equation}
for $i=c,s$, where
\begin{equation}
y_{i}=\frac{g_{i}}{\pi v_{F}},\qquad y_{4i}=\frac{g_{4i}}{\pi
v_{F}}, \label{yi}
\end{equation}
\begin{equation}
g_{i}=-g_{2\parallel}\mp g_{2\perp},\qquad g_{4i}=g_{4\parallel}\pm
g_{4\perp}, \label{gi}
\end{equation}
with the upper and lower signs in eqs. (\ref{gi}) referring to $c$
and $s$, respectively. If backscattering processes were to be
included in the theory, $g_{2\parallel} \rightarrow
g_{2\parallel}-g_{1\parallel} \equiv \tilde{g}_{2\parallel}$ in Eq.
(\ref{gi}). In addition, the $K_s$ parameter in the spin sector
would become subject to a RG flow, coupled to the marginally
irrelevant flow of $g_{1 \perp}$, the amplitude for backscattering
of electrons with opposite spins \cite{Giamarchi_book_04}. The
breaking of spin-rotational invariance by the presence of spin-orbit
interactions implies that the RG fixed-point value of $K_s$, call it
$K_s^{\ast}$, is not slaved to unity but can take larger values.
However, with the backscattering processes being weak the resulting
renormalization would be small. We will return to this issue in Sec.
V.

\subsection{Charge-, spin-, and single-particle gaps}

Given the bosonized mean-field theory defined by Eqs.
(\ref{Int_Bos_MF_Charge}) and (\ref{Int_Bos_MF_Spin}) we shall now
address the question of how electron-electron interactions influence
the Rashba-induced single-particle gap established in Sec. III.B. As
anticipated in Sec. III.C, this task gets complicated by the fact
that already for noninteracting electrons the excitation gap in the
spin-charge basis is nontrivially related to the single-particle
gap, being in effect a composite two-particle gap. Moreover, as seen
in Eq. (\ref{Delta_c-Delta_s}), the mean-field theory in the
spin-charge basis overestimates the actual size of this two-particle
gap. The situation for interacting electrons gets further confounded
by the fact that the spin and charge gaps are no longer identical,
but take on separate values, reflecting the collective nature of the
excitations in the presence of electron-electron interactions.

Taking off from Sec. III.D where we calculated the mean-field charge
soliton mass $M_c$ and spin soliton mass $M_s$ for the case of
non-interacting electrons, we perform a similar procedure, now with
electron-electron interactions included, starting with the
reparametrized sine-Gordon models in Eqs. (\ref{Int_Bos_MF_Charge})
and (\ref{Int_Bos_MF_Spin}). Note that by construction, and in exact
analogy with the noninteracting case discussed in Sec. III.D, $M_c$
and $M_s$ are the mean-field approximations of the spin and charge
gaps of the fully interacting theory, $\Delta_{c}$ and $\Delta_{s}$
respectively. Using Eqs. (\ref{Mc}), (\ref{Ms}),
(\ref{Int_Bos_MF_Charge}), and (\ref{Int_Bos_MF_Spin}), we get the
following relations between $M_c, M_s$ and $M_R$:
\begin{equation} \label{MScaling}
\eta^{-1}_{c}M_{c}=\eta^{-1}_{s}M_{s}=
\Lambda(2M_{R}/\Lambda)^{2/(4-K_c-K_s)},
\end{equation}
where $\eta_c = \eta_c(K_c, K_s)$ satisfies
\begin{multline} \label{eta_c} \eta_{c}^{16-4K_c - 4K_s} \equiv C_c^{(4-K_c)(4-K_s)} \\
\times B_c^{2K_s} C_s^{(4-K_s)K_s}B_s^{8-2K_s},
\end{multline}
with $\eta_s$ given by the same expression, but with $c
\leftrightarrow s$, and where $B_i \equiv B(K_i)$, $C_i \equiv
C(K_i), i=c,s$, are defined in Appendix A.

The mean-field version of the noninteracting theory is recovered by
choosing $K_c=K_s=1$, for which
\begin{eqnarray} \label{eta_c11}
\eta_c(1,1) &= &C_c^{9/8} B_c^{2/8} C_s^{3/8} B_s^{6/8} \mid_{K_c=K_s=1} \nonumber \\
&=& C(1)^{3/2}B(1) \nonumber \\
& \approx& 1.7,
\end{eqnarray}
with the identical number for $\eta_s(1,1)$, the result which we
arrived at already in Eq. (\ref{Delta_c-Delta_s}) via a slightly
different route. Thus, to repeat, while the mean-field theory
correctly reproduces the identity $M_c = M_s$ for noninteracting
electrons, the size of the corresponding two-particle gap  $\Delta_c
= \Delta_s$ gets overestimated by a factor of 1.7.

We can improve upon the situation by dividing away this number {\em
for all} $K_c$ and $K_s$, thus in effect defining a regularized
version of the mean-field spin and charge gaps,
\begin{equation}
\Delta_i \equiv \eta_i^{-1}(1,1)M_i, \label{Gaps}
\end{equation}
with $M_i, i=c,s$, given in Eq. (\ref{MScaling}). By construction,
this produces gives the correct noninteracting limit.

\begin{figure}[htpb]
\begin{center}
\includegraphics[scale=0.85]{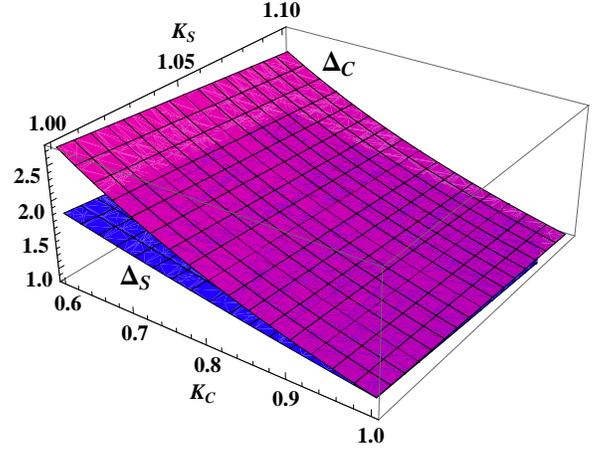}
\caption{(Color online) The mean-field regularized charge
($\Delta_c$) and spin ($\Delta_s$) gaps measured in units of the
bare gap $M_{R}$, as a function of the parameters $K_c$ and $K_s$ in
the experimentally relevant parameter range $0.6 \le K_c \le 1.0$
and $1.0 \le K_s \le 1.1$.}
\label{Fig:fig4}
\end{center}
\end{figure}

\begin{figure}[htpb]
\begin{center}
\includegraphics[scale=0.85]{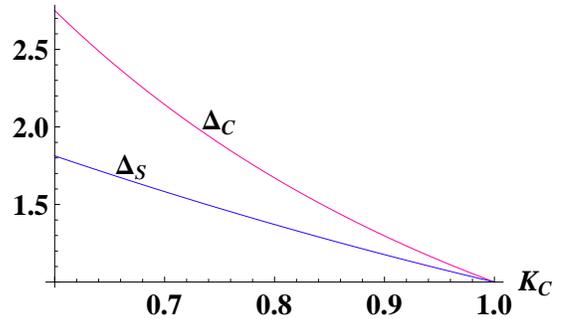}
\caption{(Color online) Projections of the mean-field regularized
charge ($\Delta_c$) and spin ($\Delta_s$) gaps on the $K_s=1.0$
plane.}
\label{Fig:fig5}
\end{center}
\end{figure}

Fig. \ref{Fig:fig4} shows $\Delta_c$ and $\Delta_s$ for the
experimentally relevant parameter range $0.6 \le K_c \le 1.0$ and
$1.0 \le K_s \le 1.1$. (As an example, to be elaborated upon in Sec.
V, a generic quantum wire obtained by gating an InAs heterostructure
is well described by taking $K_c \approx 0.7$ and $K_s \approx
1.1$.) The fundamental features of the influence of electronic
interactions on the charge and spin gaps can be gleaned from Fig.
\ref{Fig:fig5} that shows a projection of the previous surfaces on
the $K_s=1.0$ plane.

As is manifest by these curves, the important property that
$\Delta_s\leq\Delta_c$, expected on physical grounds
\cite{Giamarchi_book_04}, is respected by the regularized mean-field
gaps. The equality is valid at $K_c=K_s=1.0$, that is, in the
absence of electronic interactions, reproducing the result of Sec.
III.D.

Fig. \ref{Fig:fig5} also shows that both charge and spin gaps are
decreasing functions of the parameter $K_c$. Since $K_{c}$ decreases
with increasing $g$-couplings (c.f. eqs. (\ref{Ki})-(\ref{gi})), our
results show that the gaps increase with the $g$-couplings, i.e. are
robust against electronic interactions. Note that the impact of
electronic interactions is particularly strong on the charge gap,
which, in the considered range of parameters, more than doubles by
increasing the strength of electronic interactions.

Let us proceed to the calculation of the single-particle gap which,
in the experimentally relevant case of electron [hole] transport
through a quantum wire, defines the characteristic energy scale of
the system. In particular, this is the gap that determines the
current blockade effect, the key element in a spin transistor design
based on a modulated Rashba interaction.

Let us first recapitulate the formal definitions for single-particle
gaps following the classification used in Ref. \onlinecite{Manmana}.
The single-particle gaps are defined as the energies necessary to
add to the system one electron or one hole with spin projection $\s
\equiv\pm 1/2$:
\begin{eqnarray}
\Delta^{+}_{\s} & = & \left [ E_0(N+1,\s) - E_0(N,S=0) \right ]\,,\label{1gap+} \\
\Delta^{-}_{\s} & = & \left [ E_0(N-1,\s) - E_0(N,S=0) \right ]\,,
\label{1gap-}
\end{eqnarray}

In the case of non-interacting electrons, studied in Sec. III.B, we
identified the single-particle gaps with the mass $M_R$ of the
fermionic quasiparticles (massive sine-Gordon solitons and
antisolitons in the $\tau=\pm$ basis): $\Delta_{\tau} = M_R$.
\cite{footnoteSpin} We are now equipped to take on the calculation
of the single-particle gap in the presence of electronic
interactions. As we will not be able to resolve the particle- and
hole gaps in Eqs. (\ref{1gap+}) and (\ref{1gap-}), we instead focus
on the {\em average} single-particle gap
\begin{equation} \label{Maverage}
\bar{M} = \frac{1}{4}(\Delta^{+}_{\sigma}+
\Delta^{+}_{-\sigma}+\Delta^{+}_{\sigma^{\prime}}+
\Delta^{-}_{\sigma^{\prime}}),
\end{equation}
with $\Delta^{+}_{\sigma}+ \Delta^{+}_{-\sigma}$ corresponding to
the energy required to add two particles with opposite spin, and
$\Delta^{+}_{\sigma^{\prime}}+ \Delta^{-}_{\sigma^{\prime}}$ the
energy to add a particle and a hole with the same spin. Now, as we
saw in Sec. III.C, a charge soliton (of mass $\Delta_c$) is
precisely built from a pair of fermions carrying opposite spin, with
a spin soliton (of mass $\Delta_s$) being composed of a
particle-hole pair in a spin triplet state. While these properties
were established for the case of noninteracting electrons, the
generalizations of Eqs. (\ref{spinsoliton}) and
(\ref{chargesoliton}) to the case of rescaled fields,
\begin{eqnarray}
S_z &=&\sqrt{\frac{K_s}{2\pi}} \int_{-\infty}^{\infty} dx\, \partial_x \varphi_s(x), \\
Q &=& \sqrt{\frac{2K_c}{\pi}} \int_{-\infty}^{\infty} dx\,
\partial_x \varphi_c(x),
\end{eqnarray}
show that the relation between the gaps as determined by the
assignment of quantum numbers are unchanged by electron
interactions. It follows from Eq. (\ref{Maverage}) that
\begin{equation}
\bar{M} = \frac{1}{4}(\Delta_c + \Delta_s),
\end{equation}
from which we infer $-$ with the help of Eqs. (\ref{MScaling}) and
(\ref{Gaps}) $-$ the mean-field (average) single-particle gap
\begin{equation} \label{MMEAN}
\bar{M}_{\text{mean}} = \kappa(K_c,K_s) \Lambda\,
(2M_{R}/\Lambda)^{2/(4-K_c-K_s)},
\end{equation}
with
\begin{equation} \label{kappa}
\kappa(K_c,K_s) \equiv
\frac{1}{4}\eta_c^{-1}(1,1)(\eta_{c}(K_c,K_s)\!+\!\eta_{s}(K_c,
K_s)),
\end{equation}
and where $M_R$ is given by Eq. (\ref{MR}). Eq. (\ref{MMEAN}) is the
key result on which we shall build our analysis in the next section.
In the limiting case of non-interacting electrons, where
$K_{c}=K_{s}=1$,  $\eta_{c}=\eta_{s}$, and $\kappa(K_c,K_s)=1/2$, we
obtain, from Eq. (\ref{MMEAN}), $\bar{M}_{\text{mean}} = M_{R}$ and
thus recover the result of Sec. III.B.

\begin{figure}[htpb]
\begin{center}
\includegraphics[scale=0.65]{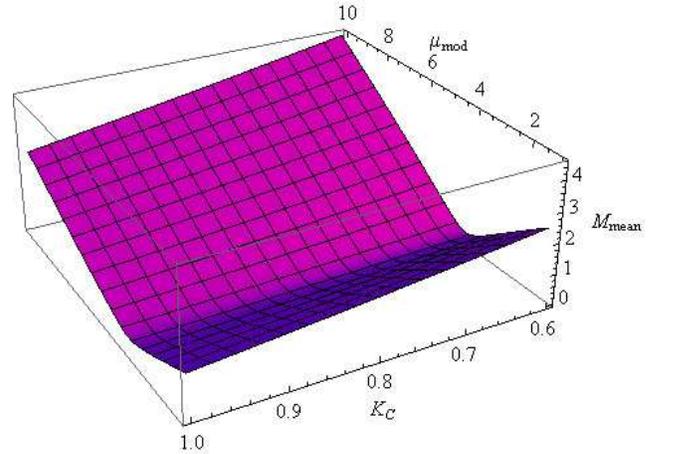}
\caption{(Color online) The mean-field value of
$\bar{M}_{\text{mean}}$ [meV] of the single-particle gap as a
function of the parameters $K_c$ and $\mu_{\text{mod}}$ in the range
$0.6 \le K_c \le 1.0$, for $K_s = 1.1$ and with $\lambda_R = - 2$
meV,  $\nu = 0.04$, $\Lambda = 100$ meV, and 1 meV $\le
\mu_{\text{mod}} \le 10$ meV.}
\label{Fig:fig6}
\end{center}
\end{figure}

In Fig. \ref{Fig:fig6} we have plotted $\bar{M}_{\text{mean}}$ in
the range $0.6\le K_c\le 1.0$, for $K_s=1.1$ and with $\lambda_R=-2$
meV, $\nu=0.04$, $\Lambda=100$ meV, and 1 meV $\le \mu_{\text{mod}}
\le 10$ meV. (The previous values correspond to the case study
carried out in Sec. V.) Again, note the significant effect of the
electron-electron interactions on the size of the single-particle
gap: decreasing the strength of electronic interactions, the gap
also decreases.

\section{Application: Towards a new type of current switch?}

In Ref. \onlinecite{JJF_Paper_09} it was argued that the insulating
gap that opens when the Rashba modulation becomes commensurate with
the band filling is sufficiently large to be exploited in a spin
transistor design. When the modulation is turned off, the electrons
are free to move and will carry a current when a drain-to-source
voltage is applied. By charging the gates $-$ thus turning on the
modulation $-$ the system becomes insulating and the current gets
blocked. A rough estimate in Ref. \onlinecite{JJF_Paper_09}, using
data for a gated InAs heterostrucuture \cite{Grundler}, suggested a
drain-to-source threshold voltage of the order of 100 mV.  In this
section we revisit the problem, now equipped with a more complete
theory which sports a refined formula for the single-particle gap,
Eq. (\ref{MMEAN}), as well as a description of effects from the
concurrent modulation of the chemical potential.

For a new type of current switch to become competitive it is
essential that the ON-OFF switching time $\tau$ compares favorable
with that of the ubiquitous MOSFETs that are used in present day
electronics. Since $\tau$ grows with the applied gate voltage (with
the power dissipation during switching being proportional to the
square of the gate voltage), we shall consider the case where the
device operates at or below a typical gate voltage of a MOSFET, in
the 1 V range or lower. While the gate-controlled built-in electric
fields in a doped heterostructure can be quite strong, the issue is
whether the resulting Rashba effect $-$ now combined with the
chemical potential modulation $-$ can become sufficiently large to
be used as a switch with a gate voltage of this moderate size. An
important constraint is here that leakage currents must be prevented
in the OFF state. The present dominance of silicon CMOS devices for
current switching is largely due to the fact that there is virtually
no leakage current in the OFF state of a MOSFET, effectively
protecting against unwanted signals as well as against standby power
dissipation. Since an assessment of possible sources of leakage
currents can only be made on basis of a specific technical design,
we will not be able to fully address this constraint here. However,
the minimal requirement that {\em thermal leakage} of charge must be
prevented will be an important bench mark in our analysis. At room
temperature, it translates into the requirement that the gap should
be $>25$ meV. Clearly, this is a lower bound. Heating of the device,
as well as the requirement that the source-to-drain voltage must not
be too small, points to a minimum gap around, maybe, 100 meV. This
is the number quoted in Ref. \onlinecite{JJF_Paper_09}, but can it
be reproduced within our more elaborate theory?

To find out, we must assign numbers to the parameters $M_R, K_c$,
$K_s$, and $\Lambda$ that enter the expression for the
single-particle gap in Eq. (\ref{MMEAN}). We shall use the same data
source as in Ref. \onlinecite{JJF_Paper_09}, obtained from the
experimental work by Grundler on gate-controlled Rashba interaction
in a square asymmetric InAs quantum well \cite{Grundler}. By
adjusting the gate voltage appropriately after application of a LED
pulse (which increases the 2D carrier density via the persistent
photo effect), Grundler succeeded to tune the Rashba spin splitting
without charging the 2D channel, thus allowing for a direct probe of
the gate-voltage dependence of the Rashba parameter $\alpha$. For
our purpose, this is an important feature, since in our model we
treat $\alpha_{\text{Rmod}}$ as being independent of the electron
density. Let us add that quantum wells based on InAs, realized in
In$_{1-x}$Ga$_{x}$As/In$_{1-x}$Al$_{x}$As\cite{Grundler,Nitta97,Hu,Koga}
or InAs/AlSb\cite{Heida} heterostructures, are preferred choices in
many proposals for spintronics applications due to their typical
large Rashba couplings $\alpha \approx 5 - 10 \times 10^{-12}$
eVm\cite{Nitta97}. Moreover, unless the quantum well is designed
with a very small valence band offset, it is also safe to assume
that $\beta \ll \alpha$, with $\beta$ the Dresselhaus
coupling\cite{Meier}. As we have found that the Dresselhaus
interaction reduces the size of the insulating gap, this is an
additional desired feature of the InAs quantum well probed in the
experiment by Grundler \cite{Grundler}. In what follows we assume
that the heterostructure studied in Ref. \onlinecite{Grundler} has
been gated so as to define a single-channel micron-range ballistic
quantum wire.

\subsection{Band gap for non-interacting electrons}

Let us start by estimating the insulating band gap for the
non-interacting theory. As already discussed, the band gap is twice
the single-fermion mass $M_R$, defined in Eq. (\ref{MR}).

Taking $\beta \ll \alpha$, we neglect the presence of the
Dresselhaus interaction completely, for which case $\lambda_R
\approx 2 \gamma_{\text{Rmod}} \sin(q_0 a)$. By inspection of Eq.
(\ref{MR}), an estimate of $M_R$ then requires numbers for
$\gamma_{\text{Rmod}}, \mu_{\text{mod}}$, $\nu$, and $q_0a$.

Beginning with $\gamma_{\text{Rmod}}$, we assume that  this
amplitude depends on the voltage of the small periodically spaced
gates that produce the modulation (see Fig. 1) in the same way as
$\gamma_R$ depends on a uniform gate voltage. This is a reasonable
assumption since $-$ neglecting random fluctuations from dopant ions
\cite{SJJ} $-$ the internal electric field in the quantum well that
supports the Rashba interaction is primarily determined by the slope
of the band edge along the growth direction of the heterostructure
(perpendicular to the 2D InAs quantum well interface), and hence its
(extremal) value right below the center of one of the small
periodically spaced gates should approach that for the case of a
single large gate. Inspection of Fig. 2 (b) in Ref.
\onlinecite{Grundler} reveals that the Rashba interaction $\alpha$
in the device considered {\em decreases} by roughly $2 \times
10^{-11}$ eVm with an increase in gate voltage of 0.1 V, indicating
that the Rashba and the chemical potential modulations are out of
phase by $\pi$. As an aside, note that the data in Fig. 2 (b) in
Ref. \onlinecite{Grundler} confirms the theoretical expectation that
the change of the Rashba coupling with applied gate voltage is {\em
linear}, a fact that we used in Sec. III.E when analyzing the
functional dependence of the effective band gap. Translating to our
geometry, the decrease of the Rashba interaction with a positive
increase of gate voltage implies that there is a minimum negative
offset, call it $- \alpha_0$, from the uniform value
$\mid\!\!\alpha_R\!\!\mid$ with no gates present. This is due to the
fact that the transverse component of the net gate electric field
(which controls the Rashba interaction) has a nonzero value at the
midpoint between two gates. In other words, the maximum value of the
{\em total} Rashba interaction (uniform + modulated) in the presence
of the small gates is given by $\alpha_R - \alpha_0$ and is attained
at the midpoint between two gates. Using the data quoted from Ref.
\onlinecite{Grundler}, it follows for the amplitude of the Rashba
modulation that $\mid\!\!\alpha_{\text{Rmod}}\!\!\mid = 1 \times
10^{-11} - \alpha_0/2$ eVm. The magnitude of the offset $\alpha_0$
is small compared to $\mid \!\!\alpha_{\text{Rmod}}\!\!\mid$ and
hence we take $\mid\!\!\alpha_{\text{Rmod}}\!\!\mid \, \approx 1
\times 10^{-11}$ eVm. We then have that
$\mid\!\!\gamma_{\text{Rmod}}\!\!\mid =
\mid\!\!\alpha_{\text{Rmod}}\!\!\mid\!/a \approx 20$ meV, where $a
\approx 5$ \AA \ is the lattice spacing in epitaxial InAs
\cite{Hornstra}. With $q_0 a \approx 0.1$, this in turn implies that
$\mid\!\!\lambda_R\!\!\mid \, \approx 2$ meV.

Turning to the amplitude $\mu_{\text{mod}}$ of the modulated
chemical potential, it is here more difficult to obtain an accurate
number and we will have to do with a rough estimate. There are two
types of contributions to $\mu_{\text{mod}}$ from the external
gates; one coming from the transverse component of the applied gate
electric field (enabling a local migration of charge from the
quantum well into the dopant layers), the other from its
longitudinal component, inducing a rearrangement of charge inside
the quantum wire. We expect the latter to dominate the modulation of
the local chemical potential and here neglect the small residual
charge migration caused by the transverse electric field. As for the
variation of the longitudinal electric field along the wire, an
exact expression requires a precise description of the fringe fields
and their superpositions from the periodic sequence of top gates.
This goes far beyond the scope of our minimal approach here, where
we assume the simplest possible, but still physically meaningful,
behavior: a longitudinal electric field that oscillates harmonically
along the wire:
$\boldsymbol{E}_{\parallel}(x)=E_{\parallel}\sin(qx)\hat{x}$,
choosing $x=0$ just below the center of one of the top gates. The
variation of the chemical potential $\Delta\mu(x)$ along the wire is
then related to the negative work required to bring an electron to
$x$,
\begin{equation}
\nonumber \Delta\mu(x)=-eE_{\parallel}\int_{0}^{x}\sin(qx')dx',
\end{equation}
\begin{equation}
\mu(x)=-\mu_{\text{mod}}\cos(qx), \label{muofx}
\end{equation}
where $\mu_{\text{mod}}=-\mu(0)=-eE_{\parallel}/q>0$. Note that the
chemical potential variation of Eq. (\ref{muofx}) is indeed the one
assumed in the very construction of our model, Eq. (\ref{mod}).

With the assumption that the amplitudes of the longitudinal and
transverse components of the net gate electric field are of the same
order of magnitude, we make the approximation $E_{\parallel} \approx
V_{g}/d$, where $V_{g}$ is the voltage of a gate electrode and $d$
is the perpendicular distance between the gate and the wire. Thus,
$\mu_{\text{mod}}=-eV_{g}/qd$. From the commensurability relation $q
= 2k_F = 2\sqrt{2\pi n_s}$, with $n_s$ the 2D electron density of
the InAs quantum well, we arrive at the desired expression for
$\mu_{\text{mod}}$ is terms of experimental parameters:
\begin{equation} \mu_{\text{mod}}=\frac{-eV_{g}}{2d\sqrt{2\pi n_s}}.
\label{mumod}
\end{equation}

From Ref. \onlinecite{Grundler} we have that $d \approx 60$ nm and
$n_s \approx 0.9 \times 10^{16}$ m$^{-2}.$ Using a gate voltage
$V_{g}\approx +0.1$ V, we obtain that $\mu_{\text{mod}} \approx 4$
meV. This number will get modified when including effects from Fermi
statistics, Coulomb interaction, and the presence of the transverse
component of the applied gate electric field. However, guided by
experimentally inferred Fermi level variations with gate voltage in
other InAs quantum wells \cite{Nitta97,Hu,Koga}, we expect that our
estimate of $\mu_{\text{mod}}$ for the geometry considered and with
the type of heterostructure used in Ref. \onlinecite{Grundler} lies
within reasonable bounds. To have some margin for error, we must
actually quote $\mu_{\text{mod}}$ only as taking possible values in
a range including 4 meV. Going back to Fig. \ref{Fig:fig4}, we see
that the single particle gap in the presence of electron-electron
interactions is not a monotonic function of $\mu_{\text{mod}}$,
displaying a minimum exactly around 4 meV. So, restricting the error
bar to lie inside the regime of interest (from the point of view of
spintronics applications) of increasing gap, we consider
$\mu_{\text{mod}}$ assuming values in the range from 4 meV to 10
meV.

In order to obtain a value for $M_R$ in Eq. (\ref{MR}) (or rather,
an interval for possible values for $M_R$, considering the numerical
uncertainty in the $\mu_{\text{mod}}$-parameter), it remains to
determine the 1D band filling $\nu$.  With the assumption that the
quantum wire has only a single conducting channel this is
straightforward. The data for the Rashba variation with top gate
voltage in Ref. \onlinecite{Grundler} were taken at a constant
electron density $n_s \approx 0.9 \times 10^{16}$ m$^{-2}$. With the
Fermi wave number $k_F$ for the quantum wire expected to be roughly
the same as for the 2D electron gas, this translates, via $k_F =
\pi\nu/a = \sqrt{2\pi n_s}$, into:
\begin{equation}
\nu=a\sqrt{\frac{2n_s}{\pi}}. \label{nu}
\end{equation}
Putting in the numbers, we get $\nu\approx 0.04$.

Inserting our estimates
\begin{displaymath}
\lambda_{\text{R}} \!\approx\!-2 \, \mbox{meV}, \ \ 4 \, \mbox{meV}
< \mu_{\text{mod}} < 10 \, \mbox{meV}, \ \ \nu \!\approx \! 0.04,
\end{displaymath}
into Eq. (\ref{MR}), we finally obtain that
\begin{equation} \label{BareMassEstimate}
0.3 \, \mbox{meV} \lesssim M_R \lesssim  3.0 \, \mbox{meV},
\end{equation}
with the upper bound corresponding to $\mu_{\text{mod}} \approx 10$
meV, and with the lower bound attained for $\mu_{\text{mod}} \approx
4$ meV. Note the minus sign in $\lambda_{\text{R}} \!\approx\! -2 \,
\mbox{meV}$, indicating that the two modulations in the type of
device considered are antiphased, which is the reason for the
non-monotonic behavior of the gap as a function of
$\mu_{\text{mod}}$ (cf. discussion in Sec. III.E).

We should here point out that given the value of $n_s$ in Ref.
\onlinecite{Grundler}, our use of a {\em single} 1D conducting
channel is not an unreasonable assumption. A first estimate, using
an infinite-well confinement potential may suggest that a quantum
wire with diameter $D\lesssim 25$ nm would satisfy the
single-channel condition: $\lambda_F = 2\pi/k_F = \sqrt{2\pi/n_s}
\approx 25$ nm. However, with the Fermi energy $E_F = \hbar^2 \pi
n_s/m^{\ast} \approx 40$ meV (with $m^{\ast}=0.04m_{e}$  for a gated
InAs quantum wire \cite{Grundler}, where $m_e$ is the electron
mass), self-consistency requires that the wire is not much wider
than roughly $D/2$ since otherwise the Fermi level would cut through
the first subband. In addition, for our modeling to make sense, our
energy cutoff $\Lambda$, first introduced in Eq. (\ref{Al-B-Z_1}),
must be smaller than the distance $\Delta E$ from the Fermi level to
the bottom of the first subband. Below we shall choose $\Lambda
\approx 100$ meV, which $-$ again assuming an infinite-well
confinement with $\Delta E = \pi^2 \hbar^2/2m^{\ast} D^2 -$ requires
the wire to be at most 7 - 8 nm wide, still, however, within the
realm of present-day technology\cite{Ferry}. One should note that a
more realistic soft confinement potential could possibly open an
additional conducting channel and lead to population of the first 1D
subband, in which case our model would no longer apply. However, as
the gate-controlled Rashba effect in Ref. \onlinecite{Grundler}
appears to be rather insensitive to a lowering of the value of
$n_s$, this potential problem should in principle be easy to
overcome (cf. Fig. 2 (b) in Ref. \onlinecite{Grundler}).

\subsection{Band gap for interacting electrons}

To complete our analysis of the single-fermion mass, we need to
include the effect of electron-electron interactions. These are
encoded by the Luttinger liquid charge- and spin parameters $K_c$
and $K_s$. Using that the forward scattering amplitudes in Eq. (56)
are all equal and given by the  $k\sim 0$ (zero momentum transfer)
Fourier component $V(k\sim 0)$ of the screened Coulomb interaction
\cite{Giamarchi_book_04}, ${g}_{2
\parallel} = g_{2 \perp} = g_{4 \parallel} = g_{4 \perp} \approx V(k
\sim 0)/\hbar$, we obtain from Eq. (56), neglecting the small
correction from backscattering processes with momentum transfer $k
\sim 2k_F$,
\begin{equation}  \label{Kc}
K_c^{-1} = \sqrt{1 + \frac{2 V(k \sim 0)}{\hbar \pi v_F}}.
\end{equation}
The screening length of the interaction is roughly set by the
perpendicular distance $d$ between the quantum wire and the nearest
metallic gate. A detailed analysis \cite{ByzcukDietl} leads to the
expression
\begin{equation} \label{Cinteraction}
V(k\sim 0) \approx \frac{e^2}{\pi \epsilon_0 \epsilon_r} \ln
(\frac{2d^{\prime}}{\xi}) + {\cal O}(\frac{\xi^2}{d^2}),
\end{equation}
where $\xi$ is the radius of the quantum wire and $\epsilon_r$ is
the averaged relative permittivity of the dopant and capping layers
between the quantum well and the nearest gate, at a distance
$d^{\prime}$ from the wire. As an interesting aside, note that the
leading logarithmic term in Eq. (\ref{Cinteraction}) depends only on
the permittivity of the environment and not on that of the wire,
implying that electrons interact mainly with image charges and not
with other electrons in the wire. With the backgate of the device in
Ref. \onlinecite{Grundler} being at a distance $d^{\prime} \approx
15$ nm from the quantum well, and with an averaged permittivity
$\epsilon_r \approx 12$ for the interjacent In$_{0.75}$Al$_{0.25}$As
and Si-doped In$_{0.75}$Al$_{0.25}$As layers \cite{Bhattacharaya},
we obtain from Eqs. (\ref{Kc}) and (\ref{Cinteraction}) that $K_c
\approx 0.7$, taking $\xi \approx$ 5 nm and using that $v_F \approx
6 \times 10^5$ m/s \cite{Bhattacharaya}. We should alert the reader
to the fact that the estimate for $K_c$ also comes with some
uncertainty, considering that it is obtained using the
parameterization in Eq. (\ref{Ki}) which is strictly valid only in
the weak-coupling limit $K_c \approx 1$. Still, {\em Bethe Ansatz}
and numerical results for this class of models have shown that the
weak-coupling formula in Eq. (\ref{Ki}) does surprisingly well in
capturing effective $K_c$ parameters also for intermediate strengths
of the electron interaction when, as in the present case, the band
filling is low, thus providing indirect support for our estimate
\cite{Schulz2}.

Turning to the spin parameter $K_s$, we already noted in Sec. IV
(text after Eq. (58)) that its bare value predicted by Eq. (56) will
renormalize to a value slightly larger than unity due to
backscattering of electrons. Having ignored these scattering
processes when writing down the spin Hamiltonian in Eq. (54) with
the rationale that they are weak in a semiconductor device, we may
compensate for their omission by adjusting the value of $K_s$ by
hand, setting it slightly larger than unity, say at $K_s \approx
1.1$. Guided by work on other models where $K_s$ takes values
different from unity  we expect this to be a reasonable estimate
\cite{GJPB_05}. In what follows $K_s$ thus represents the expected
RG fixed point value $K_s^{\ast}$, carrying an imprint of  the
marginally irrelevant backscattering term in the spin sector. It may
be worth pointing out that the correction to $K_c$ due to
backscattering is smaller than that for $K_s$ and is here neglected.

Having put numbers on $K_c$ and $K_s$ we now go back to Eq.
(\ref{eta_c}) and calculate, with the help of Eqs. (\ref{C-K}) and
(\ref{PhysicalMass}) in the Appendix,
\begin{eqnarray} \label{eta_c_number}
\eta_c(K_c \!=\! 0.7, K_s\! = \!1.1)  &\approx & 2.1 \\
\eta_s(K_c \!=\! 0.7, K_s\! = \!1.1)  &\approx & 1.4.
\end{eqnarray}

Combining this with Eqs. (\ref{eta_c11}), (\ref{MMEAN}),
(\ref{kappa}), and (\ref{BareMassEstimate}) and taking $\Lambda =
\hbar v_F/\xi \approx 100$ meV, we finally obtain an estimate for
the single-particle gap $\bar{M}_{\text{mean}}$,
\begin{equation} \label{FINAL}
0.4 \ \mbox{meV}  \lesssim \bar{M}_{\text{mean}} \lesssim 4 \
\mbox{meV}.
\end{equation}

Summarizing its specification, this result applies to a periodically
gated 5 nm thin quantum wire embedded in a Rashba-active
heterostructure of the type studied in Ref. \onlinecite{Grundler},
assuming that the Fermi wave number has been properly tuned to
commensurability with the gate spacing, and taking the gate voltage
to be +0.1 V. Note that the length scale $\xi \sim \hbar
v_F/\bar{M}_{\text{mean}}$ at which the gap starts to open up lies
within the interval 0.1 $\mu$m $\lesssim \xi \lesssim$ 1 $\mu$m,
with the lower [upper] bound corresponding to $\bar{M}_{\text{mean}}
=  4$ meV [0.4 meV], thus fitting well within the ballistic regime
of an InAs quantum wire \cite{Hornstra}.

The estimate in Eq. (\ref{FINAL}) is strikingly lower than that in
Ref. \onlinecite{JJF_Paper_09}, where the same kind of system was
analyzed using a simpler theory. One of the missing ingredients in
that theory is the interplay between the modulation of the Rashba
interaction with that of the local chemical potential, an effect
that we have found to cause a significant reduction of the gap.
Also, the gap in Ref. \onlinecite{Grundler} was extracted from that
of the collective charge excitations of the low-energy effective
model, and not, as in the present approach, properly reconstructed
as a {\em single-particle gap}, which is the one relevant for charge
transport \cite{CorrectionFootnote}.

One notes that already the gap at the upper bound in Eq.
(\ref{FINAL}) is far below the thermal threshold of $\sim 25$ meV
that is required to block thermal leakage of charge $-$ a {\em sine
qua non} for a functioning current switch. Thus, a usable device
based on spin-orbit and charge modulation effects will clearly
require a different type of heterostructure and/or input parameters
than what we have assumed here. A significant improvement would be
achieved if $-$ by ``band engineering" \cite{Litvinov,MatsudaYoh}
$-$ one could grow a heterostructure  where the  Rashba and chemical
potential modulations are {\em in phase}, not out of phase as in the
case study above. As evidenced by Eq. (\ref{MR}) and discussed in
the context of Fig. 2,  this will boost the resulting gap,
especially for the case of low electronic density. Secondly, our
analysis shows the importance of having a large electron-electron
interaction. This, in principle, is obtainable by further reducing
the electron density, with the added advantage of making the
required gate spacing larger (as seen in the commensurability
condition $q = 2k_F$ with $q=2\pi/\lambda$ and $k_F=\pi\nu/a$) thus
giving leave for larger and experimentally more tractable gates.
Finally, and most obvious, by allowing for a larger gate bias than
the 0.1 V used in the estimates above, the gap opening effect will
be further boosted. Still, unless one allows for very large
voltages, in the 5-10 V range, the gap $-$ {\em as estimated within
our formalism and for the system specifications used here} $-$ will
be too small for making a credible case for a working current switch
at room temperature. At these large voltages, however, our proposed
device would have no clear advantage compared to standard silicon
CMOS designs. Moreover, since the growth of the modulated Rashba
coupling will have saturated at much smaller voltages, any
additional gap opening effect will primarily be due to the CDW
correlations from the modulated chemical potential, not to the
presence of a Rashba interaction.

Before closing the case, however, we wish to stress that our
numerical estimates have been obtained by filtering experimental
data through an effective low-energy field theory formalism based on
a highly simplified lattice model. We have tried to be careful in
processing the data, however, the approach we use is not optimally
adapted for this task. As we have repeatedly pointed out, this makes
our numbers marred with uncertainty. Whereas our theory does provide
a ``proof-of-concept" of using a periodically gated quantum wire for
a low-bias current switch, a definite verdict about its
practicability requires more work, based on a more sophisticated
approach in modeling and analysis.

\section{Summary}

In conclusion, we have analyzed the spin- and charge dynamics in a
ballistic single-channel quantum wire in the presence of a
gate-controlled harmonically modulated Rashba spin-orbit
interaction, and with a concurrent harmonic modulation of the local
chemical potential. To be able to model a quantum wire in a gated
heterostructure with lattice inversion asymmetry, we have also
allowed for a uniform Dresselhaus spin-orbit interaction.

Depending on the relation between the common wave number $q$ of the
two modulations, the Fermi momentum $k_F$, and a parameter $q_0$
which encodes the strength of the Dresselhaus and the uniform part
of the Rashba interaction, the electrons in the wire may form a
metallic or an insulating state. Specifically, and most interesting
from the viewpoint of potential spintronics applications, when
$\mid\!\!q-2k_F\!\!\mid \ll {\cal O}(1/a)$ and
$\mid\!\!q\pm2q_0\!\!\mid \simeq {\cal O}(1/a)$ (with $a$ the
lattice spacing), a {\em nonmagnetic insulating state} is formed,
with an effective band gap which depends on the amplitudes of the
Rashba and chemical potential modulations as well as on the
strengths of the uniform Dresselhaus and Rashba interactions.
Whereas the Dresselhaus interaction {\em reduces} the band gap, the
uniform part of the Rashba interaction {\em increases} its size. The
gap also increases with the amplitude of the {\em modulated} part of
the Rashba interaction, but only if the local chemical potential
modulation is in phase with that of the Rashba interaction, {\em or}
if the amplitude of the Rashba modulation is larger than some
threshold value of the chemical potential modulation. Else, the gap
is a decreasing function of the Rashba modulation amplitude. The
resulting crossover behavior of the effective band gap is controlled
by the band filling, which sets the threshold value of the chemical
potential modulation.

This gap-opening scenario, including the crossover behavior, is
found to be robust against electron-electron interactions. To arrive
at this conclusion we used a bosonization approach, mapping the
interacting problem onto two mean-field decoupled sine-Gordon
models. By a careful analysis of the structure of the ensuing
charge- and spin gaps, we devised a regularization scheme from which
the size of the {\em single-particle gap} can be reconstructed, and
which allowed us to determine its dependence on the strength of the
electron-electron interaction. Exploiting exact results for the
sine-Gordon model we found that the gap scales as $M_R^{2/(4-K_c -
K_s)}$ where $M_R$ is the sine-Gordon soliton (or antisoliton) mass
for noninteracting electrons, and where $K_c$ and $K_s$ are the
Luttinger liquid charge- and spin parameters, respectively.  Whereas
the scaling exponent agrees with that found in Ref.
\onlinecite{JJF_Paper_09}, our estimate for the band gap (using data
from the same experimental setup as that studied in Ref.
\onlinecite{JJF_Paper_09}) comes out dramatically smaller:  As
discussed in the previous section, the theory in Ref.
\onlinecite{JJF_Paper_09} does not include the competition between
Rashba and chemical potential modulations in the experimentally
relevant parameter regime, and moreover, the band gap is not
properly reconstructed as a single-particle gap from the collective
spin- and charge excitations in the bosonic spin-charge basis
\cite{CorrectionFootnote}.

While our analysis reveals shortcomings with proposals
\cite{JJF_Paper_09,Wang,GongYang} to use present-day materials and
designs for constructing a low-bias current switch from a
gate-controlled modulated Rashba interaction, it also points to
possible routes  to overcome the problem. Besides the obvious
measure to search for materials with larger Rashba couplings, we
have shown the importance to engineer heterostructures where the
gate-controlled Rashba modulation is {\em in phase} with that of the
local chemical potential produced by the gate configuration
\cite{Litvinov,MatsudaYoh}. We have also shown that the size of the
effective band gap can be significantly boosted by reducing the
electron density in the quantum wire; this leads to a reduction of
the screening of the electron-electron interaction, and, with that,
a larger gap-opening effect from the Rashba modulation.

From a more fundamental perspective, questions about how the
gap-opening scenario is influenced by  disorder\cite{Disorder},
magnetic field effects\cite{MagneticField}, and subband mixing
\cite{NONINTERACTING} are yet to be addressed. These become
challenging problems in the context of a spatially varying
spin-orbit interaction, and may require a theoretical approach that
goes beyond the effective field-theory approach that we have used
here. Even with these questions unanswered, however, our prediction
of an electrically driven commensurate-to-incommensurate phase
transition is amenable to an experimental test. Of particular
interest would be to test for the rigidity of the insulating state
away from commensurability. As  discussed in Sec. III, its
robustness is determined by the size of the effective band gap, and
will thus be sensitive to the screening of the electron-electron
interaction, and hence to the density of electrons in the wire. By
preparing setups with different modulation wave numbers and electron
densities $-$ but otherwise identical $-$ an experiment should see
an increase of the gap with lowered electron density, as predicted
in our Eq. (\ref{MMEAN}). Our prediction that the conductivity close
to the transition scales with the chemical potential with a
universal critical  exponent 1/2 {\em independent of
electron-electron interactions}, is also open for experimental
probes.

Considering the complexity of the spin- and charge dynamics in a
quantum wire when subject to a modulated Rashba spin-orbit
interaction, further work $-$ experimental as well as theoretical
$-$ may well uncover hidden features of this fascinating physical
system.\\

\section{Acknowledgments} We wish to thank Alvaro Ferraz for valuable
discussions, and U. Ekenberg for a helpful communication. This work
was supported by the Brazilian CNPq and Ministry of Science and
Technology (M.M.), Georgian NSF Grant No. ST09/09-447 and SCOPES
Grant IZ73Z0-128058 (I.G and G.I.J), and Swedish Research Council
Grant No. 621-2008-4358 (H.J.).

\appendix
\section{Mass scales and expectation values in the sine-Gordon model}

In this appendix we show how to obtain Eqs. (\ref{Al-B-Z_1}) and
(\ref{A-B-Z_1}) from the results on the sine-Gordon model in Refs.
\onlinecite{Al_B_Zamolodchikov_95} and \onlinecite{Luk_Zam_97}.

Following the convention in Ref. \onlinecite{Luk_Zam_97}, we write
the Euclidean action of the sine-Gordon model as
\bea\label{Luk_Zam_Action} {\cal A}_{SG} &=& \int d^2
x\,\Big\{\,\frac{1}{16\pi}(\partial_{\nu}\varphi)^{2}
-\frac{2\mu_0}{a^2} \cos( \beta\varphi )\,\Big\}.
\eea
where $\mu_0$ and $\beta$ are dimensionless parameters, with $0 <
\beta^2 < 1$. (Note that the presence of the prefactor $1/16\pi$ in
the kinetic term of the action implies that the sine-Gordon coupling
$\beta$ differs by a factor of $\sqrt{8\pi}$ from the conventional
one. \cite{Coleman} Also note that by defining $\mu \equiv
\mu_0/a^2$, we have isolated the engineering dimension
1/(length)$^2$ of the bare mass $\mu$ in Ref.
\onlinecite{Luk_Zam_97} in the square of the microscopic length
$a$.) Introducing a velocity parameter $v$ via $d^2x \rightarrow v
d\tau dx$ and rescaling the field, $\varphi \rightarrow
\sqrt{8\pi}\varphi$, the corresponding Hamiltonian reads
\bea H \!= \!\!\int \!dx\Big\{{v \over 2}[(\partial_{x}\vartheta)^2
\!+\!(\partial_x \varphi)^2]\! - \frac{2\mu_0 v}{a^2}
\cos(\sqrt{8\pi} \beta \varphi)\!\Big\} \eea
with $\partial_x \vartheta$ the conjugate momentum to $\varphi$. By
the substitutions $\beta^2 = K_c/4$, $v = v_c$, and $2\mu_0 =
m_c/\pi\Lambda_c$, with $\Lambda_c = v_c/a$ a UV cutoff, we recover
the charge-sector mean-field Hamiltonian in Eq.
(\ref{Int_Bos_MF_Charge}) (when $\mu_{\text{eff}} =0$). Similarly,
the spin-sector mean-field Hamiltonian in Eq.
(\ref{Int_Bos_MF_Spin}) is obtained via the substitutions $\beta^2 =
K_s/4$, $v = v_s$, and $2\mu_0 = m_s/\pi\Lambda_s$, with $\Lambda_s
= v_s/a$, together with the phase shift $\varphi_s \rightarrow
\varphi_s + \sqrt{\pi/8K_s}$ (which does not affect the
renormalization of the theory).

The key result in Ref. \onlinecite{Al_B_Zamolodchikov_95} (encoded
in Eqs. (2.12) and (4.1) in the same reference), which relates the
sine-Gordon soliton mass to the bare mass parameter, can now be
rephrased as
\begin{equation} \label{translatedM}
M_i/\Lambda = C(K_i) (\frac{m_i}{\Lambda})^{2/(4-K_i)}, \ \ i = c,s
\end{equation}
where
\bea \label{C-K} C(K_i)\!&\!=\!&\! \frac{2}{\sqrt{\pi}}
\frac{\Gamma(\frac{K_i}{8-2K_i})}{\Gamma(\frac{2}{4-K_i})}
\left[\frac{\Gamma(1-K_i/4)}{2\Gamma(K_i/4)}\right]^{2/(4-K_i)}\!\!.
\eea
Here $\Gamma$ is the Gamma function. To extract Eqs.
(\ref{translatedM}) and (\ref{C-K}) from Eqs. (2.12) and (4.1) in
Ref. \onlinecite{Al_B_Zamolodchikov_95} we have used that
$p=(2-K_i)/2$ in these equations, and also that the sine-Gordon
action in Eq. (2.1) in Ref.  \onlinecite{Al_B_Zamolodchikov_95} is
the same as that in Eq. (\ref{Luk_Zam_Action}) after having rescaled
the field, $\varphi \rightarrow \sqrt{8\pi}\varphi$, and put $\mu_0
= \mu a^2$.

To obtain Eqs. (\ref{Al-B-Z_1}) and (\ref{A-B-Z_1}), we also need to
relate the soliton mass to the groundstate expectation value of the
cosine field; recall from Eqs. (\ref{Mc}) and (\ref{Ms}) that the
mean-field bare mass parameters $m_c$ and $m_s$ are defined in terms
of $\langle \cos\sqrt{2\pi K_s}\varphi_{s}\rangle$ and $\langle \cos
\sqrt{2\pi K_c}\varphi_{c}\rangle$, respectively. For this, we turn
to Eq. (15) in Ref. \onlinecite{Luk_Zam_97}, from which we infer
\begin{equation} \label{ExpValues}
\langle \cos(\sqrt{2\pi K_i}\varphi_i)\rangle =
B(K_i)(M/\Lambda)^{K_i/2},
\end{equation}
where
\bea \label{PhysicalMass} B(K_i)
 \!&\!=\!&\!\pi^{2} [\Gamma(1/2+\xi_i/2)\Gamma(1-\xi_i/2)]^{(K_i-4)/2} \nonumber \\
\!&\!\times\! &\!
\left[\frac{\sin(\pi\xi_i/2)}{2\sqrt{\pi}}\right]^{K_i/2}\!
\frac{(1+\xi_i) \Gamma(1-K_i/4)}{\sin(\pi \xi_i) \Gamma(K_i/4)} \eea
with $\xi_i = K_i/(4-K_i)$. Combining Eqs. (\ref{translatedM}) -
(\ref{PhysicalMass}), we obtain the following expressions for the
charge and spin soliton masses,
\bea  \frac{M_{c}}{\Lambda} &=&
C_{c}\left(\frac{2M_{R}\la\cos(\sqrt{2\pi K_s}\varphi_{s})\ra }{\Lambda}\right)^{2/\zeta_c} \nonumber \\
&=&C_{c} B_s^{2/\zeta_c}
\left(\frac{2M_{R}}{\Lambda}\right)^{2/\zeta_c}
\left(\frac{M_{s}}{\Lambda}\right)^{K_{s}/\zeta_c}, \label{Mc_Mixed}
\eea
\bea \frac{M_{s}}{\Lambda}&=&
C_{s}\left(\frac{2M_{R}\la\cos(\sqrt{2\pi K_c}\varphi_{c})\ra }{\Lambda}\right)^{2/\zeta_s} \nonumber \\
&=&C_{s} B_c^{2/\zeta_s}
\left(\frac{2M_{R}}{\Lambda}\right)^{2/\zeta_s}
\left(\frac{M_{c}}{\Lambda}\right)^{K_{c}/\zeta_s}, \label{Ms_Mixed}
 \eea
where $\zeta_i = 4-K_i, B_i = B(K_i)$, and $C_i = C(K_i), i=c, s$.
Some straightforward algebra on Eqs. (\ref{Mc_Mixed}) and
(\ref{Ms_Mixed}) finally yields Eqs. (\ref{MScaling}) and
(\ref{eta_c}). In the important limiting case of noninteracting
electrons, i.e. with $K_c=K_s=1$, we have that $B(1) \equiv {\cal
C}_1 \approx 1.0$ and $C(1) \equiv {\cal C}_0 \approx 1.4$ (cf. Sec.
III.D).


\begin{thebibliography}{99}
\bibitem{FootnoteOnComponent}
Note that the $j=1$ component in Eq. (\ref{compact}) is also present
if both the band filling and $q$ are extremely small, implying that
also $q+2k_{F} \ll {\cal O}(1/a)$. In what follows, however, we
consider the more generic situation that the commensurability
condition $|q-2k_{F}| \ll {\cal O}(1/a)$ does not hinge on the
smallness of $q$ and $k_F$, and that therefore only the $j=-1$ term
in Eq. (\ref{compact}) is present.
\bibitem{densityfootnote}
Note that the commensurability condition $q = 2k_F$ translates
precisely into $n_s = \pi/2\lambda^2$ via the relations $k_F =
\sqrt{2\pi n_s}$ and $q = 2\pi/\lambda$.
\bibitem{footnoteSpin}
Note that here, for consistency with the notation in Eqs.
(\ref{1gap+}) and (\ref{1gap-}), we use the index $\s$ instead of
$\tau$ for the spin projections along the combined uniform Rashba
and Dresselhaus fields.
\bibitem{CorrectionFootnote}
The overestimate of the effective band gap in Ref.
\onlinecite{JJF_Paper_09} is in part also due to an incorrect
parameterization of the argument of a Gamma function that enters the
expressions for the renormalized mass scales and expectation values
in the sine-Gordon model. Specifically, the expression for  ${\cal
C}_1(K_i)$ given after Eq. (13) in Ref. \onlinecite{JJF_Paper_09}
should be replaced by the right-hand side of Eq. (A4) in the present
article.
\end{thebibliography}
\end{document}